\documentclass[]{interact}

\usepackage{epstopdf}% To incorporate .eps illustrations using PDFLaTeX, etc.
\usepackage[caption=false]{subfig}% Support for small, `sub' figures and tables

\usepackage{endnotes}

\usepackage{tikz}
\usetikzlibrary{patterns,arrows,calc,decorations.pathmorphing,math}
\tikzmath{\l=50;\zs=35;\o=100;\m=100;\ls=20;\zss=45;}

\usepackage{multirow,multicol,makecell,mathtools,amsmath}

\definecolor{mygreen}{RGB}{0, 158, 115}

\DeclareMathOperator{\sgn}{sgn}

\usepackage[numbers,sort&compress]{natbib}% Citation support using natbib.sty
\bibpunct[, ]{[}{]}{,}{n}{,}{,}% Citation support using natbib.sty
\makeatletter% @ becomes a letter
\def\NAT@def@citea{\def\@citea{\NAT@separator}}% Suppress spaces between citations using natbib.sty
\makeatother% @ becomes a symbol again

\usepackage{bbold}

\newcommand* {\vek}[1]{{\bm{\mathrm{#1}}}}
\newcommand* {\rr}{\vek{r}}

\newcommand* {\Ec}{\mathcal{E}}
\newcommand* {\Bc}{\mathcal{B}}
\newcommand* {\Dc}{\mathcal{D}}
\newcommand* {\Hc}{\mathcal{H}}
\newcommand* {\Pc}{\mathcal{P}}
\newcommand* {\Mc}{\mathcal{M}}
\newcommand* {\Jc}{\mathcal{J}}

\graphicspath{{.}{./Fig/}}

\begin{document}

\articletype{REVIEW}

\title{Magnetoelectricity in two-dimensional materials}

\author{
\name{Y\`il\`e Y{\=\i}ng\textsuperscript{a,b,c} and Ulrich
Z\"ulicke\textsuperscript{a}\thanks{CONTACT Ulrich Z\"ulicke.
Email: uli.zuelicke@vuw.ac.nz}}
\affil{\textsuperscript{a}School of Chemical and Physical
Sciences, Victoria University of Wellington, Wellington, New
Zealand; \textsuperscript{b}Department of Physics and Astronomy,
University of Waterloo, Waterloo, Ontario, Canada;
\textsuperscript{c}Perimeter Institute for Theoretical Physics,
Waterloo, Ontario, Canada}
}

\maketitle

\begin{abstract}
Since the initial isolation of few-layer graphene, a plethora
of two-dimensional atomic crystals has become available,
covering almost all known materials types including metals,
semiconductors, superconductors, ferro- and antiferromagnets.
These advances have augmented the already existing variety of
two-dimensional materials that are routinely realized by
quantum confinement in bulk-semiconductor heterostructures.
This review focuses on the type of material for which
two-dimensional realizations are still being actively sought:
magnetoelectrics. We present an overview of current
theoretical expectation and experimental progress towards
fabricating low-dimensional versions of such materials that can
be magnetized by electric charges and polarized electrically
by an applied magnetic field --- unusual electromagnetic
properties that could be the basis for various useful
applications. The interplay between spatial confinement and
magnetoelectricity is illustrated using the paradigmatic
example of magnetic-monopole fields generated by electric
charges in or near magnetoelectric media. For the purpose of
this discussion, the image-charge method familiar from
electrostatics is extended to solve the boundary-value problem
for a magnetoelectric medium in the finite-width slab geometry
using image dyons, i.e., point objects having both electric
and magnetic charges. We discuss salient features of the
magnetoelectrically induced fields arising in the thin-width
limit.
\end{abstract}


\section{Introduction}

Magnetoelectric media~\cite{ode70,lan84,fie05} are characterized
by unconventional equilibrium responses to an electric field
$\vek{\Ec}$ and a magnetic field $\vek{\Bc}$. While $\vek{\Ec}$
induces only an electric polarization $\vek{\Pc}$ in ordinary
materials, it also creates a magnetization $\vek{\Mc}$ in
magnetoelectrics. Similarly atypical is an electric polarization
arising due to $\vek{\Bc}$ in addition to a magnetization. The
full range of linear electromagnetic responses occurring in a
magnetoelectric material is embodied in the constitutive
relations expressing the electric displacement field $\vek{\Dc}
\equiv\epsilon_0\, \vek{\Ec} + \vek{\Pc}$ and the field
$\vek{\Hc}\equiv\mu_0^{-1}\, \vek{\Bc} - \vek{\Mc}$ in terms of
$\vek{\Ec}$ and $\vek{\Bc}$, which read
\begin{equation}\label{eq:DHafoEB}
\vek{\Dc}=\underline{\epsilon}\, \vek{\Ec} + \underline{\alpha}
\, \vek{\Bc} \quad , \quad \vek{\Hc} = \underline{\mu}^{-1}\,
\vek{\Bc} - \underline{\alpha}^\mathrm{T}\, \vek{\Ec} \quad .
\end{equation}
Here $\underline{\epsilon}$ and $\underline{\mu}$ are the
familiar~\cite{jac99} materials tensors of electric permittivity
and magnetic permeability, respectively. The magnetoelectric
tensor $\underline{\alpha}$ is associated with the
magnetic-field-induced electric polarization, and its
transpose $\underline{\alpha}^\mathrm{T}$ governs the
electric-field-induced magnetization~\cite{ode70,lan84,fie05}.
The interested reader can find a more in-depth discussion of
constitutive relations for magnetoelectrics in
Appendix~\ref{app:ConstRel}, including a juxtaposition of two
commonly adopted conventions.

Seminal works on the magnetoelectric effect were largely focused
on the symmetry classification of materials with nonvanishing
$\underline{\alpha}$~\cite{dzy59,sch73}, but ramifications of
magnetoelectric responses have also been discussed early
on~\cite{sir94}. However, it was the advent of multiferroic
materials~\cite{fie16,spa17} with their large magnetoelectric
couplings~\cite{eer06,don15,spa19,don19} that has greatly
boosted current interest in gaining a deeper understanding
about magnetoelectricity~\cite{hu17}, also with the view to
enable useful applications~\cite{chu07,fus14,ort15,son17,
mani19}. The intriguing connection with the particle-physics
concept of axion electrodynamics~\cite{sik83,wil87,heh08a} has
recently become even more relevant through the realization of
topological insulators~\cite{qi08,ess09,arm19,nen20,sek21} and
Weyl semimetals~\cite{nen20,sek21,ma15,den21}. The opportunity
to explore sizable magnetoelectric effects in real materials has
spurred efforts to describe unconventional electromagnetic
phenomena exhibited by magnetoelectrics~\cite{qi09,fec14,mei19,
mar16,mar21,oue19,och12,kho14,kho21,uri20,kam20}. Among these,
simulations of magnetic-monopole fields have been
paradigmatic~\cite{qi09,fec14,mei19,kho14,uri20}.

Nanostructuring generally opens up new avenues for controlling
materials properties, including those of multiferroics and
magnetoelectrics~\cite{vel11,hu15}. Recent advances in the
design of two-dimensional (2D) atomic crystals~\cite{nov05} and
their versatile combination~\cite{nov16,liu16,sie21} have
transcended the original platform of low-dimensional systems
created in semiconductor heterostructures~\cite{bau84,dav98}.
With magnetic 2D materials already realized in
metal~\cite{him99} and semiconductor quantum wells~\cite{die14}
as well as atomic crystals~\cite{gib19,gon19a,mak19,wei20}, the
focus has shifted to the possibility of 2D multiferroics and
magnetoelectrics~\cite{gon19a,mak19,wei20,lu19,tan19,zho20,
gao21a}. In this context, the purpose of this Review is
two-fold. In Sec.~\ref{sec:matSur}, we present a survey of 2D
materials where magnetoelectricity is expected and/or has
already been observed. In the subsequent Sec.~\ref{sec:2DEB}, we
discuss the ramification of a 2D material's finite size on its
magnetoelectric responses, focusing especially on
magnetic-monopole fields generated by electric charges placed
near or in the medium. Our work is intended to raise awareness
of the special features associated with magnetoelectricity in
low-dimensional materials, which have been discussed only
sporadically until now~\cite{pou21}.

\section{Survey of 2D magnetoelectric materials}
\label{sec:matSur}

The systematic discussion of magnetoelectric materials
properties is usually based on consideration of a medium's
free-energy density $F(\vek{\Ec}, \vek{\Bc})$. Its expansion
up to quadratic order in electric and magnetic field components
is most generally given by~\cite{lan84,fie05}
\begin{equation}\label{eq:freeEB}
F(\vek{\Ec}, \vek{\Bc}) = F(\vek{0}, \vek{0}) -
\vek{\Pc}_\mathrm{s}\cdot \vek{\Ec} - \vek{\Mc}_\mathrm{s}\cdot
\vek{\Bc} - \frac{1}{2}\, \vek{\Ec}\cdot \big(
\underline{\chi}_\Ec\, \vek{\Ec}\big) - \frac{1}{2}\, \vek{\Bc}
\cdot \big( \underline{\chi}_\Bc\, \vek{\Bc}\big) - \vek{\Ec}
\cdot \big(\underline{\alpha}\, \vek{\Bc}\big) - \dots \, .
\end{equation}
Here $\vek{\Pc}_\mathrm{s}$ and $\vek{\Mc}_\mathrm{s}$ are the
material's spontaneous electric polarization and magnetization,
respectively, while $\underline{\chi}_\Ec$ and
$\underline{\chi}_\Bc$ denote the conventional~\cite{jac99}
electric and magnetic susceptibility tensors. The linear
magnetoelectric effect is embodied in the term containing
$\underline{\alpha}$. Contributions to $F(\vek{\Ec}, \vek{\Bc})$
giving rise to nonlinear electromagnetic (including higher-order
magnetoelectric~\cite{fie05,asc68,gri94}) responses have been
omitted from Eq.~(\ref{eq:freeEB}), as we are not focusing on
these in the following. Similarly, possible interactions between
simultaneously present spontaneous electric and magnetic orders,
such as $\vek{\Mc}_\mathrm{s}$ being affected when switching
$\vek{\Pc}_\mathrm{s}$ in multiferroics, are not being
considered in the following unless these present a mechanism
for generating a novanishing $\underline{\alpha}$.

General descriptions of materials properties in terms of tensor
quantities can be obtained based on fundamental symmetry
considerations~\cite{nye57}. For example, magnetoelectricity can
occur only in systems where both spatial-inversion and
time-reversal symmetries are broken. This allows us to
distinguish the magnetoelectric responses  we are interested in
from the superficially related phenomena of current-induced
magnetization~\cite{gan19} or spin-orbit torque~\cite{man19},
which are based on a relation $\vek{\Mc} = \underline{\eta}\,
\vek{\Jc}$ between the magnetization and an electric-current
density $\vek{\Jc}$ ~\cite{ivc78,bel78,aro91,ede90}. If the
current originates from an applied electric field and is related
to the latter via Ohm's law, $\vek{\Jc} = \underline{\sigma}\,
\vek{\Ec}$, the resulting dependence of the magnetization on the
electric field,
\begin{equation}
\vek{\Mc} = \underline{\eta}\, \vek{\Jc} = \underline{\eta}\,
\big( \underline{\sigma}\, \vek{\Ec} \big) \equiv
\underline{\eta}^\prime\, \vek{\Ec} \,\, ,
\end{equation}
looks similar to the magnetoelectric effect~\cite{lev85}; see
Eq.~(\ref{eq:MfromE}). However, the basic conditions for the
tensor $\underline{\eta}$ to exist in a material differ from
those required for nonvanishing $\underline{\alpha}$. Most
importantly, current-induced magnetization is possible in
materials with unbroken time-reversal
symmetry. Furthermore, in contrast to the magnetoelectric
effect which is an equilibrium phenomenon, nonequilibrium
mechanisms are crucial for generating a magnetization
accompanying the electric current~\cite{gan19}. Recent
theoretical studies have explored current-induced magnetization
in 2D materials, e.g., in twisted bilayer graphene~\cite{he20}
and the interfacial 2D electron gas in SrTiO$_3$~\cite{joh21}.
Spin-orbit torques in magnetic 2D materials such as
Fe$_3$GeTe$_2$~\cite{joh19} and CrI$_3$~\cite{xue21} have also
been considered. An intriguing realization of current-induced
magnetization in the 2D transition-metal dichalcogenide
MoS$_2$~\cite{lee17} utilized the material's valley-isospin
degree of freedom~\cite{xu14}.

We focus here only on \emph{magnetoelectric\/} 2D materials,
i.e., those having finite $\underline{\alpha}$. To impose some
order on the rapidly expanding zoo of 2D magnetoelectrics, 
materials are grouped into subsections according to the basic
mechanisms underlying their magnetoelectric responses. We start
by discussing 2D versions of single-phase magnetoelectrics and
multiferroics in Sec.~\ref{sec:2Dsingle}. Following that, 
Sec.~\ref{sec:2Dmultiferro} provides an overview of 2D
multiferroic heterostructures. Sec.~\ref{sec:tune2Dmag} is
dedicated to magnetic 2D materials whose (ferro- or
antiferro-)magnetic order is tunable by electric fields. The
unconventional magnetoelectricity of 2D charge carriers in
semiconductor quantum wells is considered in
Sec.~\ref{sec:qWellME}. We conclude the survey of 2D
magnetoelectric materials with Sec.~\ref{sec:graphene} exploring
the ways in which the valley-degree of freedom enables
magnetoelectric responses in 2D atomic crystals such as
few-layer sheets of graphene or transition-metal
dichalcogenides.

\subsection{Single-phase 2D magnetoelectrics}
\label{sec:2Dsingle}

Before the advent of multiferroics, the typical magnetoelectric
bulk material was expected to be an inversion-asymmetric
time-reversal-breaking antiferromagnet, much like the first and
paradigmatic example Cr$_2$O$_3$~\cite{dzy59,rad62}. Although
time-reversal symmetry is always broken in ferromagnets but not
in all antiferromagnets~\cite{tin64}, inversion symmetry is
rarely broken in ferromagnets unless the underlying crystal
structure is already noncentrosymmetric. In contrast,
antiferromagnetic order itself can break inversion symmetry,
thus greatly multiplying the possibilities for realizing a
magnetoelectric material. By now, many antiferromagnetic
compounds are known to be magnetoelectric, but the magnitude of
their magnetoelectric couplings $\alpha_{ij}$ are usually
small~\cite{fie05} --- the current record holder is TbPO$_4$
with $\big|\alpha_{ij}\big|\lesssim 0.220\,\sqrt{\epsilon_0/
\mu_0}$~\cite{rad84,riv09}.

Despite the fact that the smallness of magnetoelectric couplings
is generic for single-phase antiferromagnets, those among them
that are van-der-Waals crystals can serve as an attractive
starting point for fabricating 2D magnetoelectrics. Pertinent
examples are MnPS$_3$~\cite{res10a} and NiI$_2$~\cite{kur13},
whose intralayer antiferromagnetic order has been shown to
persist in the few-layer limit~\cite{chu20,lon20,ju21}. To date,
there has been no direct confirmation of the magnetoelectric
effect in these 2D allotropes of bulk magnetoelectrics, but the
presence of all necessary ingredients is promising and should
motivate further exploration of these materials~\cite{ni21} and
similar candidates~\cite{chi16,mcg20}.

An interesting alternative to planar antiferromagnets is
presented by layered magnets where the magnetization direction
alternates between adjacent layers. Bilayers of such materials
could constitute single-phase antiferromagnets and show the
magnetoelectric effect. A prominent example is CrI$_3$ for which
the magnetoelectric coupling has been measured to be
$|\alpha_{zz}| \sim 0.034\,
\sqrt{\epsilon_0/\mu_0}$~\cite{jia18}. See also related
experiments~\cite{hua18,jia18a} and theoretical
work~\cite{siv18,lei21}. It can be expected that similar
materials such as WSe$_2$ and CrTe$_2$ also exhbibit
magnetoelectricity~\cite{lei21}.

Single-phase materials with large magnetoelectric effects are
typically multiferroic, i.e., exhibit coexisting spontaneous
electric polarizations and magnetizations. There is, however, a
paucity of such materials even in bulk, because the most
effective mechanisms for generating ferromagnetisms and
ferroelectricity are mutually exclusive~\cite{hil00} due to a
classic case of chemical contra-indication~\cite{spa20}: 
ferroelectrics favor having empty \textit{d} or \textit{f}
orbitals, whereas magnetism typically arises when these orbitals
are partially filled. In fact, no 2D materials that are
single-phase ferromagnetic-ferrolectric multiferroics have been
realized so far~\cite{gao21a}, but theoretical studies are
pointing to a number of promising candidates: CrN and
CrB$_2$~\cite{luo17}, MXene bilayers~\cite{li17}, monolayer
(CrBr$_3$)$_2$Li~\cite{hua18a}, TMPCs-CuMP$_2$X$_6$~\cite{qi18,
lai19}, monolayer Hf$_2$VC$_2$F$_2$~\cite{zha18}, monolayer
VOCl$_2$~\cite{ai19} and VOI$_2$~\cite{tan19a} (but see also
Ref.~\cite{din20}), halogen-intercalated phosphorene
bilayers~\cite{yan17}, transition-metal-intercalated MoS$_2$
bilayers~\cite{tu19}, double-perovskite bilayers~\cite{zha20},
as well as monolayer $\alpha$-In$_2$Se$_3$~\cite{duan21}.

\subsection{2D multiferroic heterostructures}
\label{sec:2Dmultiferro}

Hybrid systems incorporating ferroelectric and ferromagnetic
parts provide a promising avenue for overcoming the intrinsic
limitations that make it so hard to have both a spontaneous
electric polarization and a magnetization in the same
material~\cite{fie16}. In fact, gigantic magnetoelectric
couplings of $|\alpha_{ij}| \sim 4.8\times 10^3\,
\sqrt{\epsilon_0/\mu_0}$ have recently been realized in the
heterostructure multiferroic FeRh/BTO~\cite{che14}. The
expectation that a similar strategy of combining 2D
ferroelectrics with 2D ferromagnets will yield robust 2D
multiferroics has fuelled intense theoretical efforts. We
discuss below a number of promising materials combinations for
which magnetoelectic phenomena in the wider sense have been
predicted. However, the available theoretical results do not
permit us to infer with certainty that the \emph{linear\/}
magnetoelectric effect occurs, or estimate the magnitude of
magnetoelectric-tensor components $|\alpha_{ij}|$.

The first-considered van-der-Waals heterostructure of
ferromagnetic Cr$_2$Ge$_2$Te$_6$ combined with ferroelectric
In$_2$Se$_3$ showed intriguing properties~\cite{gon19}. Besides
demonstrating the ability to switch the magnetization of the
Cr$_2$Ge$_2$Te$_6$ layer via reversal of the electric
polarization in the In$_2$Se$_3$ part, the latter became
magnetized via the proximity effect --- in effect becoming a
single-phase 2D multiferroic. 

The combination of magnetic bilayer CrI$_3$ with ferroelectric
monolayer Sc$_2$CO$_2$ also yielded interesting magnetoelectric
responses~\cite{lu20}. Here the reversal of electric
polarization in Sc$_2$CO$_2$ triggered the transition between
antiferromagnetic and ferromagnetic order in bilayer CrI$_3$.
The same mechanism for magnetoelectricity has also been studied
theoretically for a bilayer-CrI$_3$/monolayer-In$_2$Se$_3$
multiferroic heterostructure~\cite{yan21}. Even greater
efficiency for electric control of the
antiferromagnetic-to-ferromagnetic transition in bilayer CrI$_3$
is expected when it forms a van-der-Waals heterostructure with
perovskite-oxide ferroelectrics such as BiFeO$_3$~\cite{li21}.

Control of the magnetization in Fe$_n$GeTe$_2$ layers via the
electric polarization of In$_2$Se$_3$ was predicted to enable
enhanced magnetotransport functionality in a tunnel-junction
device~\cite{su21}. Transition-metal-decorated graphene can also
be used as the magnetic part of a multiferroic heterostructure
formed with monolayer In$_2$Se$_3$~\cite{sha21}. 

\subsection{2D materials with electrically tunable
spontaneous magnetization}\label{sec:tune2Dmag}

Magnetic order generally arises from the exchange interaction
between microscopic magnetic moments. In systems where the
exchange-interaction strength depends on tunable electronic
degrees of freedom, e.g., the charge density in a
semiconductor~\cite{die14}, the spontaneous magnetization
$\vek{\Mc}_\mathrm{s}$ can become a function of the applied
electric field $\vek{\Ec}$. Electric control of magnetism based
on such mechanisms is attracting a lot of interest~\cite{mat15}
but has typically not been discussed using the language and
formalism of the magnetoelectric effect. Nevertheless, in the
spirit of the expansion (\ref{eq:freeEB}) for the
field-dependent part of the system's free-energy density, a term
$$-\vek{\Mc}_\mathrm{s}(\vek{\Ec})\cdot \vek{\Bc}\equiv
-\vek{\Mc}_\mathrm{s}(\vek{0})\cdot \vek{\Bc} - \left( \left.
\frac{\partial\vek{\Mc}_\mathrm{s}}{\partial\vek{\Ec}}
\right|_{\vek{\Ec}\to\vek{0}}\vek{\Ec}\right)\cdot \vek{\Bc}
-\dots$$ actually incorporates magnetoelectricity with a tensor
$\underline{\alpha}$ whose elements $\alpha_{ij} = \big[
\partial \Mc_{\mathrm{s}j}(\vek{\Ec})/\partial\Ec_i
\big]_{\vek{\Ec}\to \vek{0}}$ quantify the electric-field
tunability of the spontaneous magnetization. Typical magnitudes
$|\alpha_{ij}|\sim 4.0\times 10^{-3}\,\sqrt{\epsilon_0/\mu_0}$
have been realized in the dilute magnetic semiconductor
GaMnAs~\cite{saw10}.

Magnetoelectric effects arising from an electric-field-dependent
spontaneous magnetization have been discussed
theoretically~\cite{lee02,lee02a} and observed
experimentally~\cite{bou02,anh15} in 2D quantum-well structures
made from dilute-magnetic semiconductors. Few-layer
Cr$_2$Ge$_2$Te$_6$ has recently been shown to exhibit
electric-field-tunable ferromagnetism~\cite{xin17,wan18,sun19}.
Using experimental data and results of calculations presented in
Ref.~\cite{wan18}, we estimate $|\alpha_{ij}| \sim 2\times
10^{-6}\, \sqrt{\epsilon_0/\mu_0}$ for this 2D magnetoelectric
material. It would be interesting to explore whether other
currently available electrically tunable 2D magnets, e.g.,
monolayer Fe$_2$GeTe$_2$~\cite{den18}, also exhibit the linear
magnetoelectric effect.

\subsection{Magnetoelectricity of 2D charge carriers in
quantum wells}\label{sec:qWellME}

In bulk conductors, screening and nonequilibrium responses
associated with the application of an electric field or the
generation of an electric polarization generally preclude any
discussion of the magnetoelectric effect.\endnote{Despite
being sometimes loosely linked with magnetoelectricity, the
current-induced magnetization~\cite{gan19} that is possible in
gyrotropic conductors and the chiral magnetic effect exhibited
by Weyl semimetals~\cite{ma15,den21} are fundamentally different
from the equilibrium-magnetoelectric responses focused on here.}
Low-dimensional conductors, however, can exhibit equilibrium
responses as long as $\vek{\Ec}$ and $\vek{\Pc}$ are parallel to
spatial directions within which charge-carrier motion is
quantized~\cite{bas83,cas07,du21}. Furthermore, 2D-conductor
samples fitted with front and back gates allow changing of
$\vek{\Ec}$ at constant charge density~\cite{pap99,hab04,zha09,
shi15}, thus making it possible to disentangle
electric-polarization responses from charging effects in
experiments. Conducting 2D materials are therefore ideally
suited for a detailed exploration of magnetoelectricity
associated with itinerant charge carriers. In this Subsection,
we focus on the paradigmatic example of 2D conductors realized
in semiconductor-heterostructure quantum wells~\cite{bau84}.
A pedagogical introduction to the basic principles of band-gap
engineering and physical properties of 2D semiconductor
structures is available, e.g., in  Ref.~\cite{dav98} (especially
Chapter 4).

\begin{figure}
\centering
\subfloat[Electric-field-induced magnetization in quantum wells.
Left panel: Quadrupolar equilibrium currents (indicated in red)
are associated with antiferromagnetic order of charge carriers.
An electric field applied perpendicular to the quantum well
distorts these currents so that a dipolar current distribution
results, thus generating an in-plane magnetization.]{%
\resizebox*{6.9cm}{!}{\includegraphics{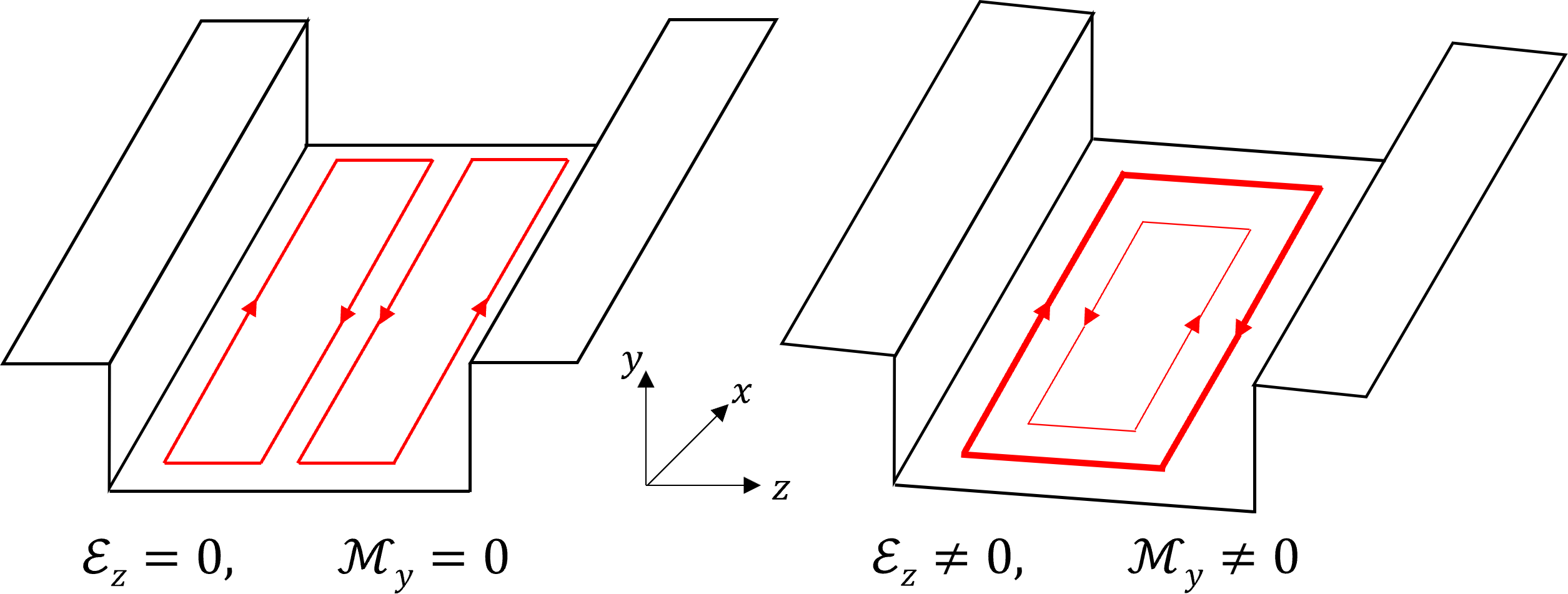}}}
\hspace{10pt}
\subfloat[Magnetic-field-induced electric polarization in
quantum wells. In the presence of antiferromagnetic order, an
applied in-plane magnetic field couples to the quantum-well
bound-state charge distribution (inciated in red), resulting in
an asymmetric shift. The latter represents an electric
polarization perpendicular to the quantum well.]{%
\resizebox*{6.9cm}{!}{\includegraphics{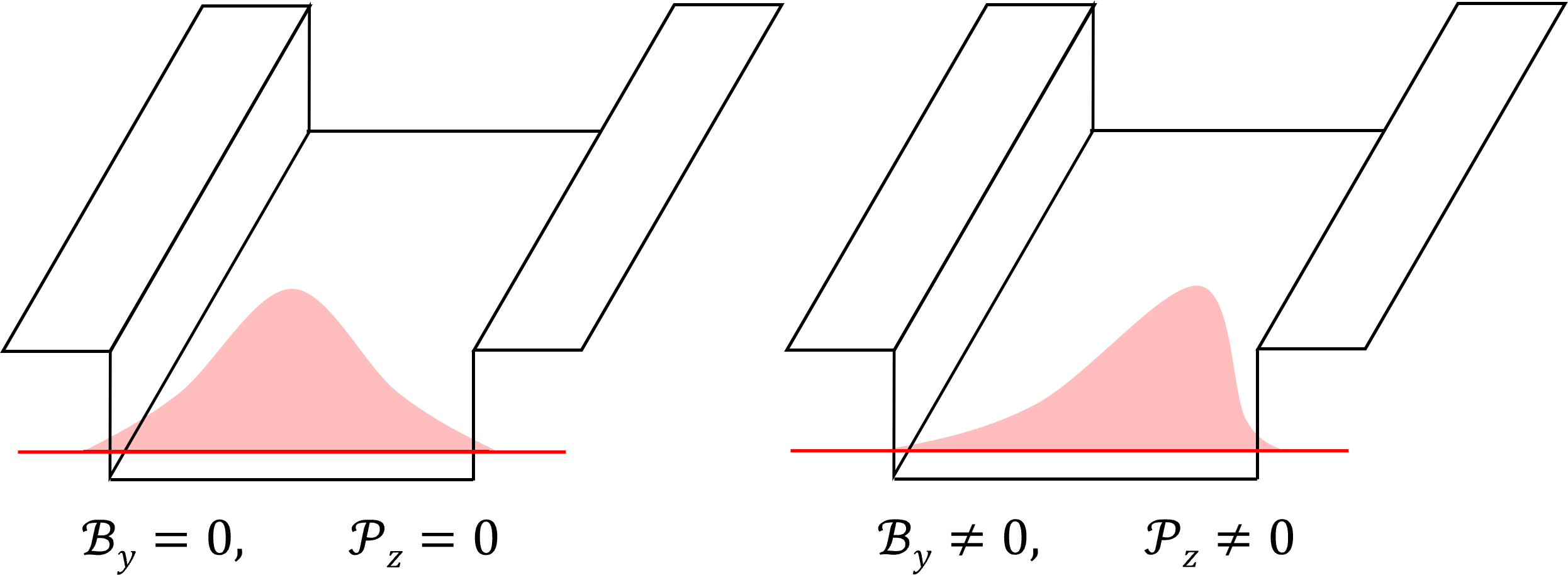}}}
\caption{\label{fig:2DME}%
Mechanisms underpinning magnetoelectricity in
semiconductor-heterostructure quantum wells. Charge carriers can
move freely in the $x y$ plane but are confined in $z$ direction
by a symmetric potential arising from the spatially varying
conduction-band bottom (drawn in black pen). Equilibrium
currents [(a)] and bound-state charge distributions [(b)] are
indicated in red. The band bending shown schematically in the
right panel of subfigure (a) is due to the applied perpendicular
electric field. The qualitatively similar, but typically much
smaller in magnitude, band bending arising as a consequence of
the magnetic-field-induced electric polarization is not shown in
the right panel of subfigure (b).}
\end{figure}

The microscopic origin of magnetoelectricity in quantum wells
are quadrupolar equilibrium currents whose distortion by an
applied electric field generates a magnetization~\cite{win20}.
Such equilibrium-current distributions arise from the interplay
of spatial-inversion asymmetry with an applied magnetic
field~\cite{gor94,win20}, or with an intrinsic exchange field
due to ferromagnetic order~\cite{gor94,win20}, or with the
staggered exchange field of a nontrivial antiferromagnetic
order~\cite{win20}.\endnote{The first scenario where the
application of a magnetic field causes the quadrupolar
equilibrium currents that are the basis for magnetoelectricity
technically constitutes a field-induced (i.e., higher-order)
magnetoelectric effect~\cite{win20,asc68}. Note also that the
mechanism for magnetoelectricity to occur via equilibrium
quadrupolar currents in a ferromagnetic-semiconductor quantum
well is fundamentally different from the magnetoelectric
response arising from the charge-density tunability of the
spontaneous magnetization discussed in
Sec.~\ref{sec:tune2Dmag}.} Furthermore, an applied magnetic
field shifts the 2D-bound-state charge distribution, thus
inducing an electric polarization perpendicular to the
well~\cite{gor94,win20}, as mandated by the duality of
magnetoelectric reponses~\cite{ode70,lan84}. 
Figure~\ref{fig:2DME} illustrates schematically the microscopic
mechanisms for magnetoelectric responses in
semiconductor-heterostructure quantum wells. According to
theoretical estimates~\cite{win20}, the magnitude of the
magnetoelectric-tensor components in quantum wells can reach
values $|\alpha_{ij}|\sim 10^{-4}\, \sqrt{\epsilon_0/\mu_0}$,
and the $\vek{\Ec}$-induced magnetic moment per 2D charge
carrier can be as large as $0.6$ Bohr magnetons.

\subsection{Magnetoelectricity in 2D materials from topology or
valley isospin}\label{sec:graphene}

Exotic types of magnetoelectricity have been suggested to exist
in certain materials, even those in which time-reversal and
spatial-inversion symmetries are not broken. One example for
these are topological insulators~\cite{qi08,ess09,arm19,nen20,
sek21}, which have an isotropic magnetoelectric coupling
$\underline{\alpha} = \alpha \,\mathbb{1}$ with a quantized
magnitude $|\alpha| \equiv e^2/(4\pi\hbar) \approx (1/137)
\sqrt{\epsilon_0/\mu_0}$. Among the multitude of 2D materials, a
topological magnetoelectric effect was predicted~\cite{zha19,
otr19,li19}, and its signatures have recently been
observed~\cite{liu20,gao21}, in even-layer
MnBi$_2$Te$_4$.\endnote{To be precise, even-layer MnBi$_2$Te$_4$
is actually antiferromagnetic and therefore not a
time-reversal-invariant topological insulator but an axion
insulator~\cite{sek21}. This distinction matters because the
fate of axion electrodynamics in the 2D limit of nonmagnetic
topological-insulator slab geometries is rather
intricate~\cite{arm19}.} While interesting from a fundamental
point of view, the inability to adjust the magnitude of the
topological magnetoelectric coupling limits its relevance for
applications --- unless proposals for engineering tunable
magnetoelectric couplings via symmetry-breaking finite-size
effects in thin MnBi$_2$Te$_4$ layers~\cite{zhu21} come to
fruition.

Materials with a valley-isospin degree of freedom~\cite{xu14}
lend themselves to an intriguing realization of electrically
tunable magnetoelectricity. Concrete proposals for 2D materials
have been formulated for bilayer graphene~\cite{zue14,kam19} and
transition-metal-dichalcogenide bilayers~\cite{gon13}. In all
these cases, electronic states from different valleys in the
2D-material band structure (usually labelled $\vek{K}$ and
$\vek{K'}$) are connected via time reversal or spatial
inversion. An imbalance between charge-carrier densities in the
two valleys --- a finite valley-isospin density --- amounts to a
breaking of both symmetries~\cite{du21} and renders the material
magnetoelectric. As various ways for addressing valley isospin
have been explored as part of a drive to realize electronic
devices based on a new valleytronics paradigm~\cite{sch16},
these systems offer the unique opportunity for generating
electrically tailored magnetoelectric couplings.

\section{Electromagnetism of 2D magnetoelectric media}
\label{sec:2DEB}

The survey of 2D magnetoelectric media presented in the previous
section outlines the great progress being made in designing,
fabricating and characterizing such materials. Once the remaining
materials-science challenges are overcome, the focus of further
investigations will shift to exploring more broadly how such
materials behave electromagnetically. In anticipation of that,
this section focuses on a paradigmatic property of
magnetoelectric media: the simulation of magnetic-monopole fields
discussed extensively for bulk systems~\cite{qi09,fec14,mei19,
kho14}. Below we briefly describe how isotropic magnetoelectric
responses can be included as unconventional source terms into the
Maxwell equations of ordinary electrodynamics, thus mimicking the
effects of axion electrodynamics~\cite{sik83,wil87,heh08a,qi08}.
This forms the basis for solving boundary-value problems
involving isotropic magnetoelectrics and enables a generalization
of the well-known method of images~\cite{jac99} to include image
magnetic monopoles alongside image charges~\cite{qi09}. In the
following, we refer to this approach as the method of \emph{image
dyons} and use it to obtain electric and magnetic fields
generated by an electric charge placed near or in a finite-width
magnetoelectric slab.

A configuration of external sources (i.e., electric charges and
currents) determines the electromagnetic fields via the
inhomogeneous Maxwell equations~\cite{jac99}
\begin{equation}\label{eq:MaxInhom}
\vek{\nabla}\cdot\vek{\Dc} = \rho \quad , \quad
\vek{\nabla}\times\vek{\Hc}  - \frac{\partial\vek{\Dc}}{\partial
t} = \vek{\Jc} \quad .
\end{equation}
Here $\rho(\rr)$ and $\vek{\Jc}(\rr)$ denote the charge and
current densities, respectively, as a function of position $\rr
\equiv (x, y, z)$, and $\vek{\nabla}\equiv (\partial/\partial x,
\partial/\partial y, \partial/\partial z)$ is the gradient
operator. Solving (\ref{eq:MaxInhom}), in conjunction with the
set of homogeneous Maxwell equations
\begin{equation}
\vek{\nabla}\cdot\vek{\Bc} = 0 \quad , \quad
\vek{\nabla}\times\vek{\Ec} + \frac{\partial
\vek{\Bc}}{\partial t} = 0
\end{equation}
and the constitutive relations (\ref{eq:DHafoEB}) that embody
the effect of microscopic degrees of freedom in media, yields
the fields throughout space.

As a starting point for exploring the interplay between low
dimensionality and magnetoelectricity, we focus in the following
on the case of media with isotropic magnetoelectric response
whose magnetoelectric tensor has the form $\underline{\alpha} =
\alpha\, \mathbb{1}$. For consistency, isotropy is also assumed
for the ordinary polarization and magnetization responses, i.e.,
$\underline{\epsilon} = \epsilon\, \mathbb{1}$ and
$\underline{\mu} = \mu\, \mathbb{1}$. To make modifications
arising from isotropic magnetoelectricity more explicit, the
field-source relations (\ref{eq:MaxInhom}) can be rewritten as
\begin{equation}\label{eq:axMax}
\vek{\nabla}\cdot\left( \epsilon\, \vek{\Ec}\right) = \rho -
\left( \vek{\nabla} \alpha \right) \cdot \vek{\Bc} \,\, , \,\,
\vek{\nabla}\times\left( \mu^{-1}\, \vek{\Bc} \right) -
\frac{\partial\left( \epsilon\, \vek{\Ec}\right)}{\partial t} =
\vek{\Jc} + \left( \vek{\nabla} \alpha \right) \times \vek{\Ec}
+ \frac{\partial\alpha}{\partial t}\, \vek{\Bc} \,\, .
\end{equation}
The particular form of the inhomogeneous Maxwell equations given
in (\ref{eq:axMax}) coincides with the one derived in the
context of axion electrodynamics~\cite{sik83,wil87} whose
realization in condensed-matter systems is currently attracting
great interest~\cite{nen20,sek21}. It shows that, in situations
with piecewise-constant $\alpha$, e.g., when certain parts of
space are filled with a homogeneous isotropic magnetoelectric
medium, the effects of magnetoelectricity can be represented in
terms of charges and currents induced at interfaces.

Various theoretical techniques have been used to solve the set
of equations describing electromagnetisms with magnetoelectric
media~\cite{qi09,fec14,mar16,mei19,oue19,mar21}. Below we
discuss in greater detail the method of introducing image
charges and monopoles~\cite{qi09} to satisfy boundary conditions
of electromagnetic fields at an interface between vacuum and an
isotropic magnetoelectric medium. Generalizing this
approach to the case with more than a single interface enables
us to investigate how the finite width of a 2D magnetoelectric
medium affects its unconventional electromagnetism.

\subsection{Image dyons for solving boundary-value problems in
magnetoelectrics}

The method of image charges is an established technique for
solving boundary-value problems in electrostatics~\cite{jac99}.
It is based on the possibility to use an appropriately placed
fictitious (i.e., image) point charge to represent the
polarization-charge distribution generated at an interface by
the presence of a real electric point charge. This approach
turns out to be most useful when considering high-symmetry
configurations, e.g., when the real charges are placed in spaces
with planar, cylindrical or spherical interfaces between
different media. In particular, the image-charge method has been
applied to obtain the electric fields generated by a charge
near~\cite{som00} or inside~\cite{kum89} a dielectric plate.

\begin{figure}
\centering
\subfloat[Image dyons for an electric charge $q_0$ above a
semi-infinite magnetoelectric medium.]{%
\resizebox*{6cm}{!}{\includegraphics{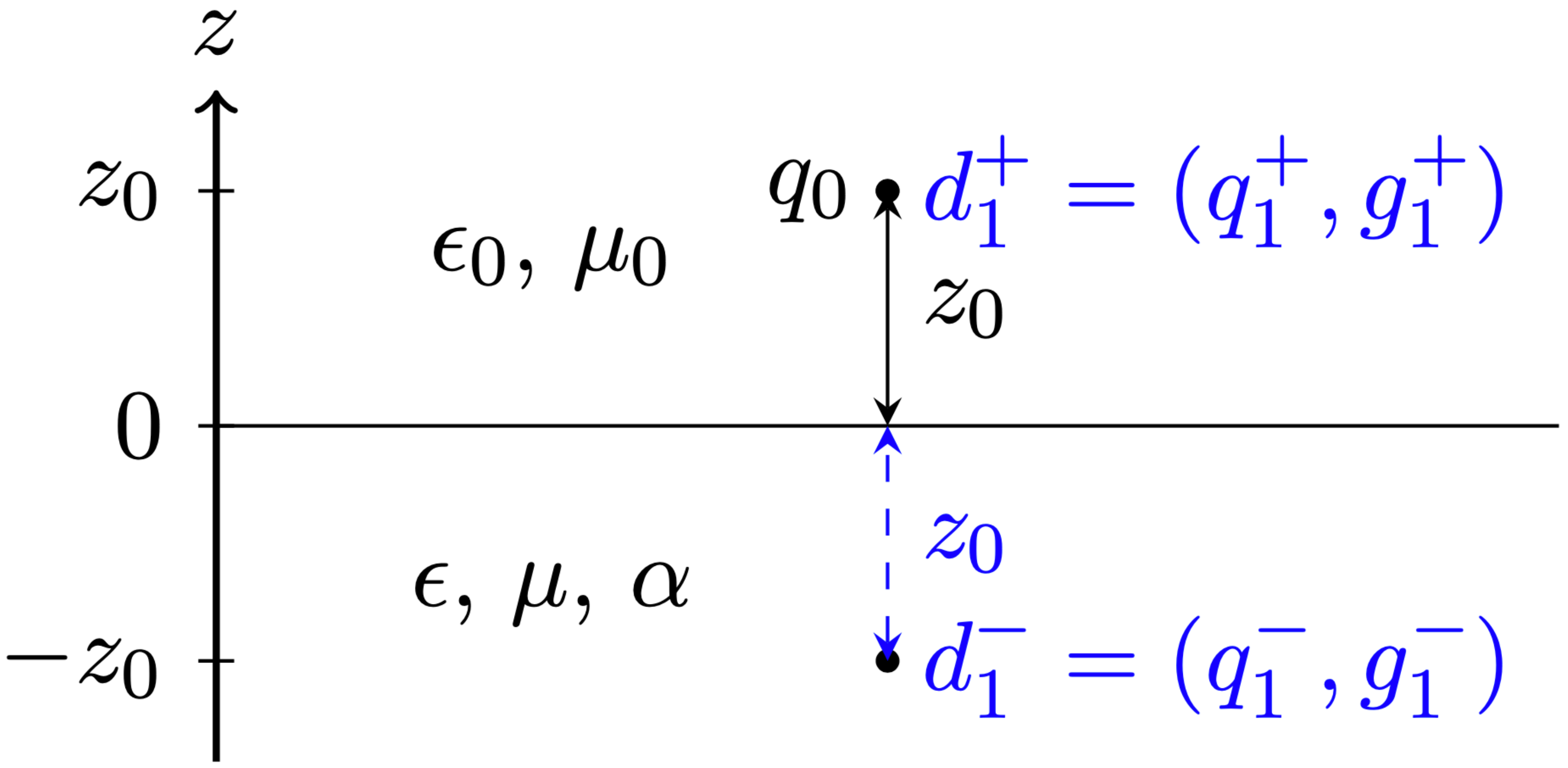}}}
\hspace{20pt}
\subfloat[Image dyons for an electric charge $q_0$ inside a
semi-infinite magnetoelectric medium.]{%
\resizebox*{6cm}{!}{\includegraphics{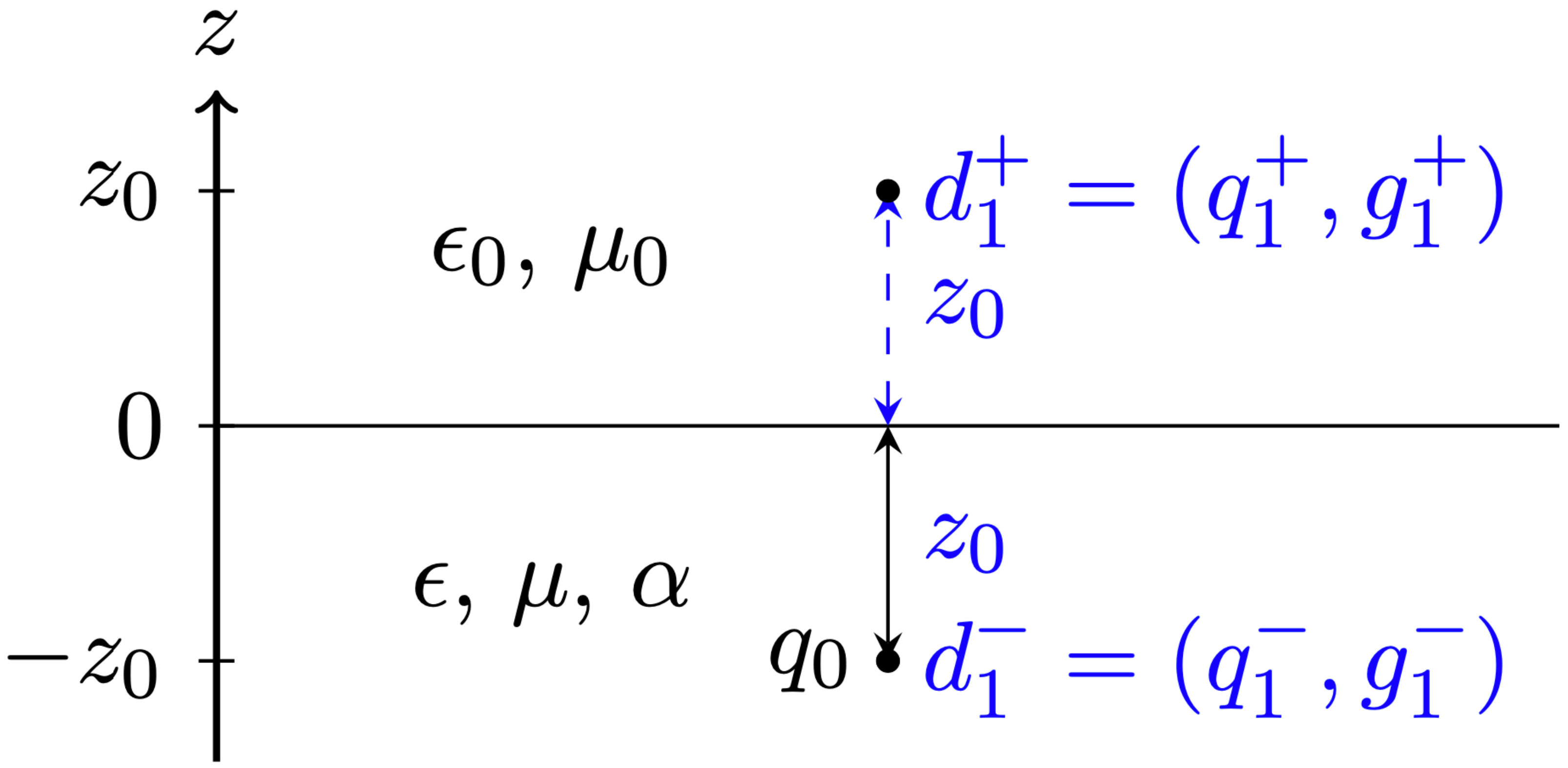}}}
\caption{\label{fig:qSemi}%
An electric charge $q_0$ placed near a planar interface (located
at $z=0$) between an ordinary dielectric (occupying the region
$z>0$) and a magnetoelectric medium (present in region $z<0$)
generates image dyons on both sides. The values of their
electric (magnetic) charges are related via $q^{+}_1=q^{-}_1$
($g^{+}_1=-g^{-}_1$).}
\end{figure}

\begin{figure}
\centering
\subfloat[Image dyons for an electric charge $q_0$ outside of a
magnetoelectric slab of width $w$.]{%
\resizebox*{6.1cm}{!}{\includegraphics{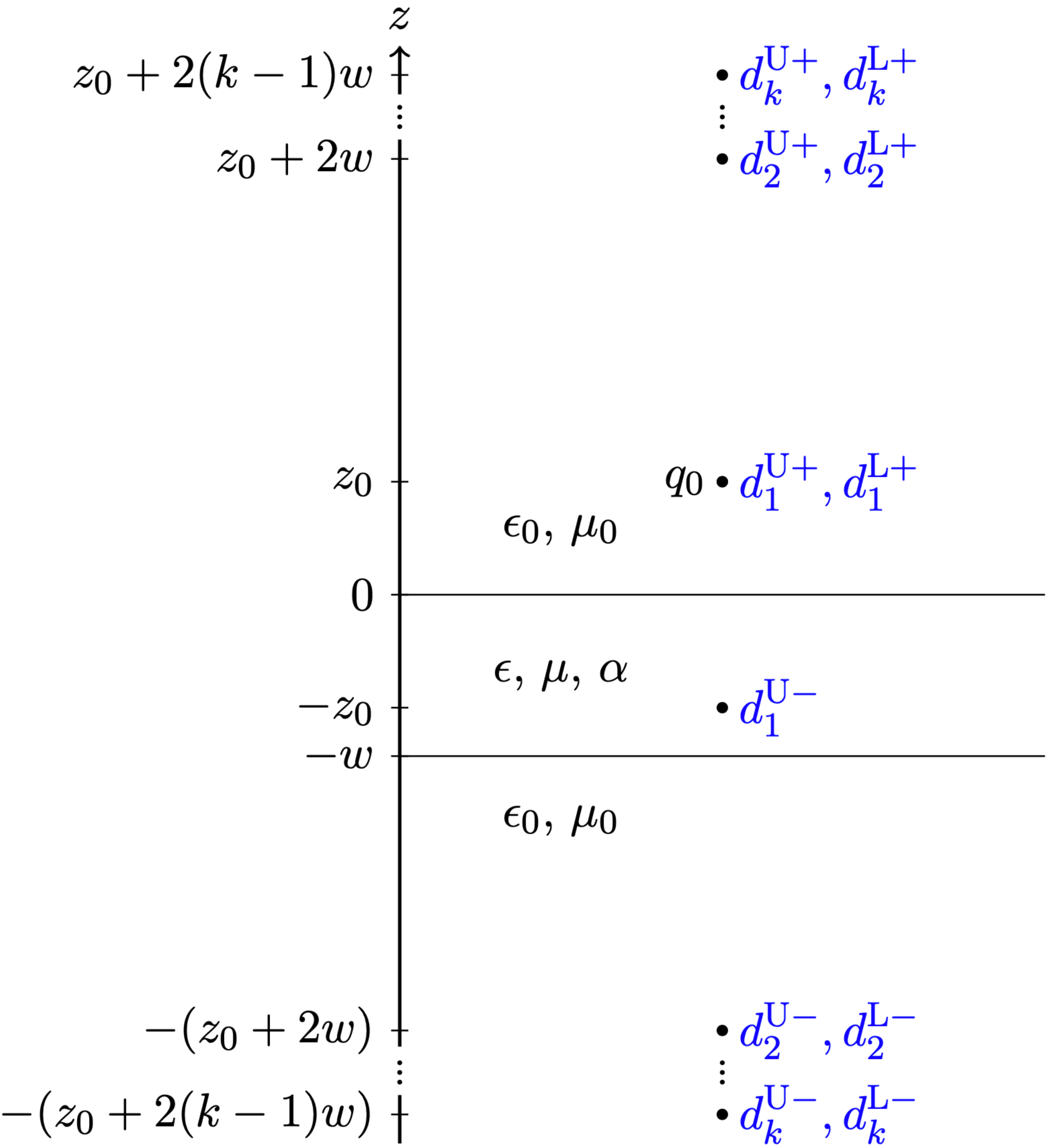}}}
\hspace{15pt}
\subfloat[Image dyons for an electric charge $q_0$ inside a
magnetoelectric slab of width $w$.]{%
\resizebox*{7.5cm}{!}{\includegraphics{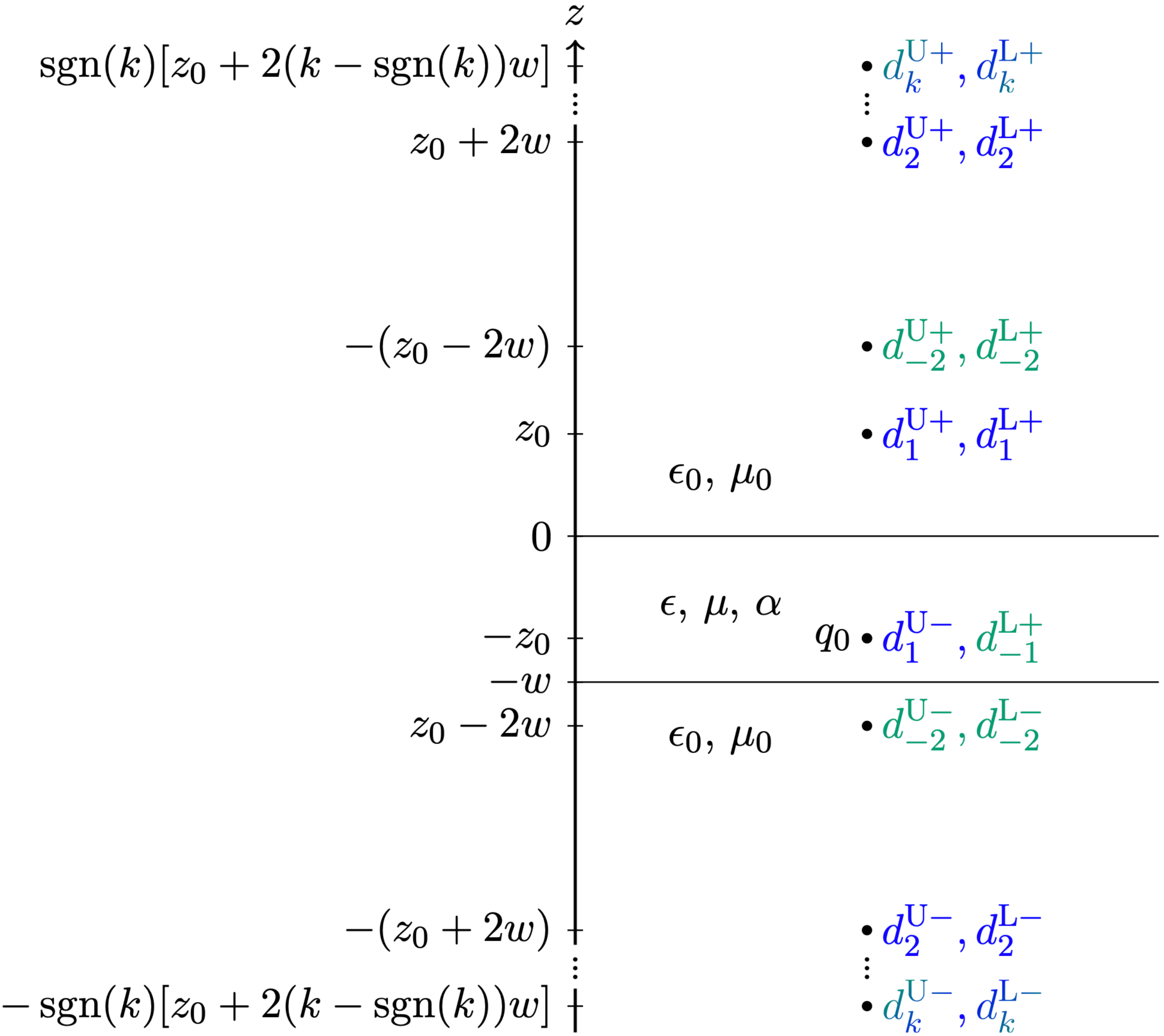}}}
\caption{\label{fig:Qslab}%
An infinite set of image dyons $d^{m\nu}_{k}$ is needed to
describe electric and magnetic fields generated by an electric
point charge $q_0$ placed near [panel (a)] or inside [panel (b)]
a finite-width slab of magnetoelectric material. Here $m=\rm{U\,
(L)}$ labels dyons arising from mirror-imaging at the upper
(lower) boundary, $\nu=+\, (-)$ when the image dyon is located
above (below) the $m$ boundary, and $k\in\mathbb{Z}\setminus 0$
is a counter associated with the dyon location. Dyons with $k<0$
arise only for the case with $q_0$ inside the magnetoelectric
medium [panel (b)].}
\end{figure}

The method of image charges can be extended to describe
situations involving magnetoelectrics by allowing magnetic
monopoles~\cite{dir31} alongside each of the electric image
charges~\cite{qi09}, thus effectively rendering the image
objects to be dyons~\cite{sch69,wit79}. See Fig.~\ref{fig:qSemi}
for an illustration of the previously considered case of an
electric charge near a single planar boundary between vacuum and
a magnetoelectric medium~\cite{qi09,mei19,pla21}. Below we
juxtapose the known results for this case with those calculated
for our situation of interest, which is a charge placed near or
inside a finite-width slab of magnetoelectric material.
Analogous to the situation of an electric point charge near or
inside a dielectric plate where an infinite number of image
charges need to be considered~\cite{som00,kum89}, an infinite
number of image dyons are required for describing the
finite-slab magnetoelectric. Figure~\ref{fig:Qslab} shows the
configurational details pertaining to this case. Before
presenting our results, we briefly state the boundary-value
problem posed by planar interfaces with magnetoelectric media
and sketch the formalism underpinning the image-dyon method used
for its solution.

Considering the static limit and assuming any field sources
entering the r.h.s.\ of the inhomogeneous Maxwell equations
(\ref{eq:MaxInhom}) to be localized away from a planar interface
that is parallel to the $x y$ plane and intersects the $z$ axis
at $z=z_\mathrm{b}$, the electromagnetic fields obey the
boundary conditions
\begin{subequations}\label{eq:BC}
\begin{eqnarray}
&& \Dc_z(R, \theta, z_\mathrm{b}^+) = \Dc_z(R, \theta,
   z_\mathrm{b}^-)\,\, , \,\, \Hc_R(R, \theta, z_\mathrm{b}^+) =
   \Hc_R(R, \theta, z_\mathrm{b}^-)\,\, , \\
&& \Ec_R(R, \theta, z_\mathrm{b}^+) = \Ec_R(R, \theta,
   z_\mathrm{b}^-)\,\, , \,\, \Bc_z(R, \theta, z_\mathrm{b}^+) =
   \Bc_z(R, \theta, z_\mathrm{b}^-)\,\, .
\end{eqnarray}
\end{subequations}
Here cylindrical coordinates $\rr\equiv (R, \theta, z)$ are
utilized for positions, and $z_\mathrm{b}^\pm =
\lim_{\varepsilon\to 0} (z_\mathrm{b} \pm \varepsilon)$.
Furthermore, the static limit enables both the electric field
and the magnetic field to be expressed in terms of scalar
potentials,
\begin{equation}
\vek{\Ec} = -\nabla V\quad , \quad \vek{\Bc} = -\nabla U \quad .
\end{equation}
The method of image dyons amounts to writing
\textit{Ans\"atze\/} for the scalar potentials arising in a
given region of space, due to having a source charge $q_0$ at
the location $\rr_0$, in terms of a superposition of
electric-point-charge and magnetic-monopole contributions,
\begin{equation}\label{eq:Ansatz}
V(\rr) = \frac{1}{4\pi\epsilon'}\left[\frac{q_0}{|\rr - \rr_0|}
+ \sum_{k} \frac{q_k}{|\rr - \rr_k|}\right] \,\, , \,\,
U(\rr) = \frac{\mu'}{4\pi} \sum_{k} \frac{g_k}{|\rr - \rr_k|}
\,\, .
\end{equation}
Here  $q_k$ ($g_k$) denote the electric (magnetic) charge of the
fictitious image dyon $d_k$ located at $\rr_k$. It is understood
that the sum $\sum_k$ is restricted to include only those image
dyons that are relevant for calculating the fields in a given
region, i.e., those that are located outside. Furthermore,
$\epsilon'$ ($\mu'$) is the permittivity (permeability) of the
region where the source charge is located. To begin with, the
electric and magnetic charges making up each image dyon $d_k
\equiv (q_k ,g_k)$ as well as its location $\rr_k$ are unknown.
However, the location $\rr_k$ can be predicted based on the
geometry of the system if we require it to be independent of the
charges $q_k$ and $g_k$. It turns out that $\rr_k$ is the
location of first- or higher-order mirror images of the source
charge with respect to the boundary (or boundaries). Since the
slab geometry considered in this work has planes perpendicular
to the $z$ axis as its boundaries, and since we assume that the
electric source charge is located at $\rr_0 = (0, 0, \pm z_0)$
with $z_0 > 0$, the locations of the image dyons are $r_k=(0, 0,
z_k)$ with the coordinate $z_k$ determined by $z_0$ and the slab
width $w$. In particular, if the magnetoelectric medium is
semi-infinite (i.e., there is only one boundary, assumed to be
located at $z=0$ without loss of generality), we only have two
image dyons $d^{+}_1 = (q^{+}_1, g^{+}_1)$ and $d^{-}_1 =
(q^{-}_1, g^{-}_1)$ as shown in Fig.~\ref{fig:qSemi}, and
Eq.~\eqref{eq:Ansatz} simplifies. Furthermore, we find the
relations $q^{+}_1 = q^{-}_1$ and $g^{+}_1 = -g^{-}_1$. Full
details of the calculation and results for image-dyon charges
are given in Appendix~\ref{app:semiinf}.

If the magnetoelectric medium has the shape of a finite-width
slab, an infinite number of image dyons exists as shown in
Fig.~\ref{fig:Qslab}. We can see this by trying to satisfy the
field-continuity conditions for each of the two boundaries in
turn. Assuming the source charge $q_0$ to be outside the slab,
we can start by satisfying the upper boundary conditions while
ignoring the lower boundary --- this yields the image dyons
$d^{\rm{U+}}_{1}=(q^{\rm{U+}}_{1},g^{\rm{U+}}_{1})$ at $z=z_0$
and $d^{\rm{U-}}_{1}=(q^{\rm{U-}}_{1},g^{\rm{U-}}_{1})$ at
$z=-z_0$ as in the single-interface situation depicted in
Fig.~\ref{fig:qSemi}(a). [Here and in the following, the
superscript U (L) is used to indicate that the image dyon
results from mirroring at the upper (lower) boundary, which is
at $z=0$ ($z=-w$), the $+$ ($-$) label next to it specifies that
the charge is located above (below) that boundary, and the
integer $k$ relates to the image-dyon location as explained
below.] Attempting next to satisfy the boundary conditions at
the bottom interface while ignoring the upper boundary, we treat
the dyon $(q_0+q^{\rm{U+}}_{1},g ^{\rm{U+}}_{1})$ at $z_0$ as
the source and obtain additional image dyons $d^{\rm{L+}}_{1}$
at $z=z_0$ and $d^{\rm{L-}}_{2}$ at $z=-(z_0+2w)$. However,
their introduction causes boundary conditions at the upper
interface to no longer be satisfied. To remedy this, two more
image dyons need to be created; $d^{\rm{U+}}_{2}$ at $z=z_0+2w$
and $d^{\rm{U-}}_{2}$ at $z=-(z_0+2w)$, whose properties are
determined from assuming dyon $d^{\rm{L-}}_{2}$ as the source.
This in turn causes boundary conditions at the lower interface
to be violated, which needs to be fixed by further iteration of
the image-dyon method. In the end, an infinite number of image
dyons are needed to satisfy the boundary conditions at both
interfaces asymptotically. 

For the case when the source charge $q_0$ is inside the slab,
even more image dyons are needed as shown in
Fig.~\ref{fig:Qslab}(b). Similar as before, we can start by
satisfying the upper boundary conditions while ignoring the
lower boundary, and obtain image dyons $d^{\rm{U+}}_{1}=$ at
$z=z_0$ and $d^{\rm{U-}}_{1}$ at $z=-z_0$. Next, we satisfy the
lower boundary conditions while ignoring the upper boundary and
obtain $d^{\rm{L+}}_{-1}$ at $z=z_0$, $d^{\rm{L-}}_{-2}$ at $z=
z_0-2w$, $d^{\rm{L+}}_{1}$ at $z=z_0$, and $d^{\rm{L-}}_{2}$ at
$z=-(z_0+2w)$ by treating both $q_0$ and the dyon
$d^{\rm{U+}}_{1}$ as sources. The subset of these image dyons
having $k>0$ are equivalent to the image dyons found for the
case with the source charge outside the slab, and further
iteration builds up the same structure as shown in
Fig.~\ref{fig:Qslab}(a). In contrast, the subset of image dyons
labelled by $k<0$ are entirely new, and their subsequent
alternate mirroring creates infinitely many additional dyons.

Generally, the location of an image dyon $d^{m\nu}_{k}$ is at
\begin{equation}
\rr^{\nu}_{k} = (0, 0, \nu\sgn(k)\{z_0 + 2 [k - \sgn(k)]w\})
\quad .
\end{equation}
Similar to the case where the magnetoelectric medium is
semi-infinite, we have $q^{m+}_{k}=q^{m-}_{k}$ and $g^{m+}_{k} =
-g^{m-}_{k}$. The general formulae for calculating the
image-dyon charges in the above mentioned iterating process are
derived in Appendix~\ref{app:dyonsource}, and the results are
given in Appendix~\ref{app:finite}. Note that the relevant image
dyons for determining electromagnetic fields in different
regions are different. The fields present in the part of space
$z>0$ ($z<-w$) are equivalent to those created by the source
charge and image dyons $d^{m+}_{k}$ ($d^{m-}_{k}$). The field
inside the medium is equivalent to that created by the source
charge and the image dyons $d^{\mathrm{U}+}_{k}$ and
$d^{\mathrm{L}-}_{k}$. By superposing the fields generated by
the source charge and all relevant image dyons and taking the
asymptotic limit, we can obtain the field configurations in
different regions.

The general principles of the image-dyon method and its detailed
results can also be formally obtained by rewriting
Eq.~\eqref{eq:Ansatz} using the geometry-adapted Sommerfeld-type
identity~\cite{som49,che99,cai13}
\begin{equation}\label{eq:interpret}
\frac{1}{|\rr -\rr_k|} = \frac{2}{\pi}\, \int^{\infty}_0
\int^{\infty}_0 d\gamma \, d\eta\,\, \frac{\cos\left[ \gamma (x
- x_k) \right]\, \cos\left[ \eta (y - y_k) \right]}{u}\,\,
\mathrm{e}^{-u\, |z - z_k|}\,\, ,
\end{equation}
where $u\equiv \sqrt{\gamma^2+\eta^2}$. Due to cylindrical
symmetry with respect to the $z$ axis, we can focus on
calculating the scalar potentials in one of the planes including
$z$ axis and then extend the results to other planes by
rotation. Thus assuming $\rr = (R, 0, z)$ in cylindrical
coordinates, and for $\rr_0 = (0, 0, z_0)$, we find
\begin{subequations}\label{eq:AnsatzSlab}
\begin{eqnarray}
V(\rr) &=& \frac{1}{2\pi^2 \epsilon'} \int^{\infty}_0
\int^{\infty}_0 d\gamma\, d\eta\,\, \frac{\cos(\gamma R)}{u}\,
\left[ q_0\, \mathrm{e}^{-u |z-z_0|} + C_1\, \mathrm{e}^{u z} +
C_2\, \mathrm{e}^{-u z} \right] , \\
U(\rr) &=& \frac{\mu'}{2\pi^2} \int^{\infty}_0 \int^{\infty}_0
d\gamma \, d\eta\,\, \frac{\cos(\gamma R)}{u} \left[ C_3\,
\mathrm{e}^{u z} + C_4\, \mathrm{e}^{-u z}\right] .
\end{eqnarray}
\end{subequations}
After determining the initially unknown coefficients $C_j$ via
the boundary conditions, $C_{1,2}$ ($C_{3,4}$) can be
interpreted in terms of electric charges $q_k$ (magnetic charges
$g_k$) and their locations $\rr_k$. Furthermore, $C_1=C_3=0$
($C_2=C_4=0$) in the region above (below) the slab since we
require scalar potentials to vanish at infinity. Thus the
image-dyon representation is obtained via
Eq.~\eqref{eq:interpret}.

\subsection{Fields from a point charge placed in or near a
magnetoelectric plate}

\begin{figure}
\centering
\subfloat[Electric point charge located outside a semi-infinite
magnetoelectric medium.]{%
\resizebox*{4.5cm}{!}{\includegraphics{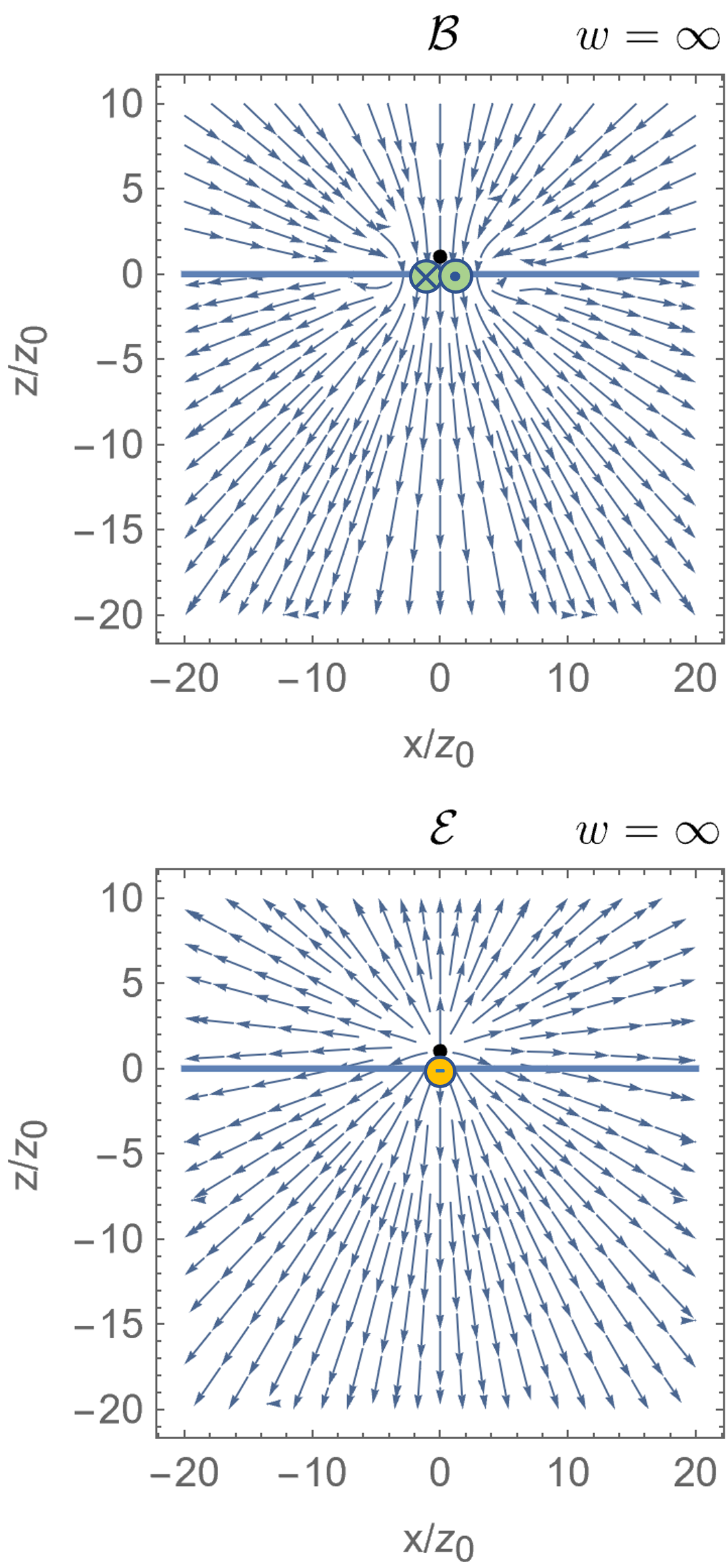}}}
\hspace{3pt}
\subfloat[Electric point charge located near a thick slab of
magnetoelectric material, i.e., $w\gg z_0$.]{%
\resizebox*{4.5cm}{!}{\includegraphics{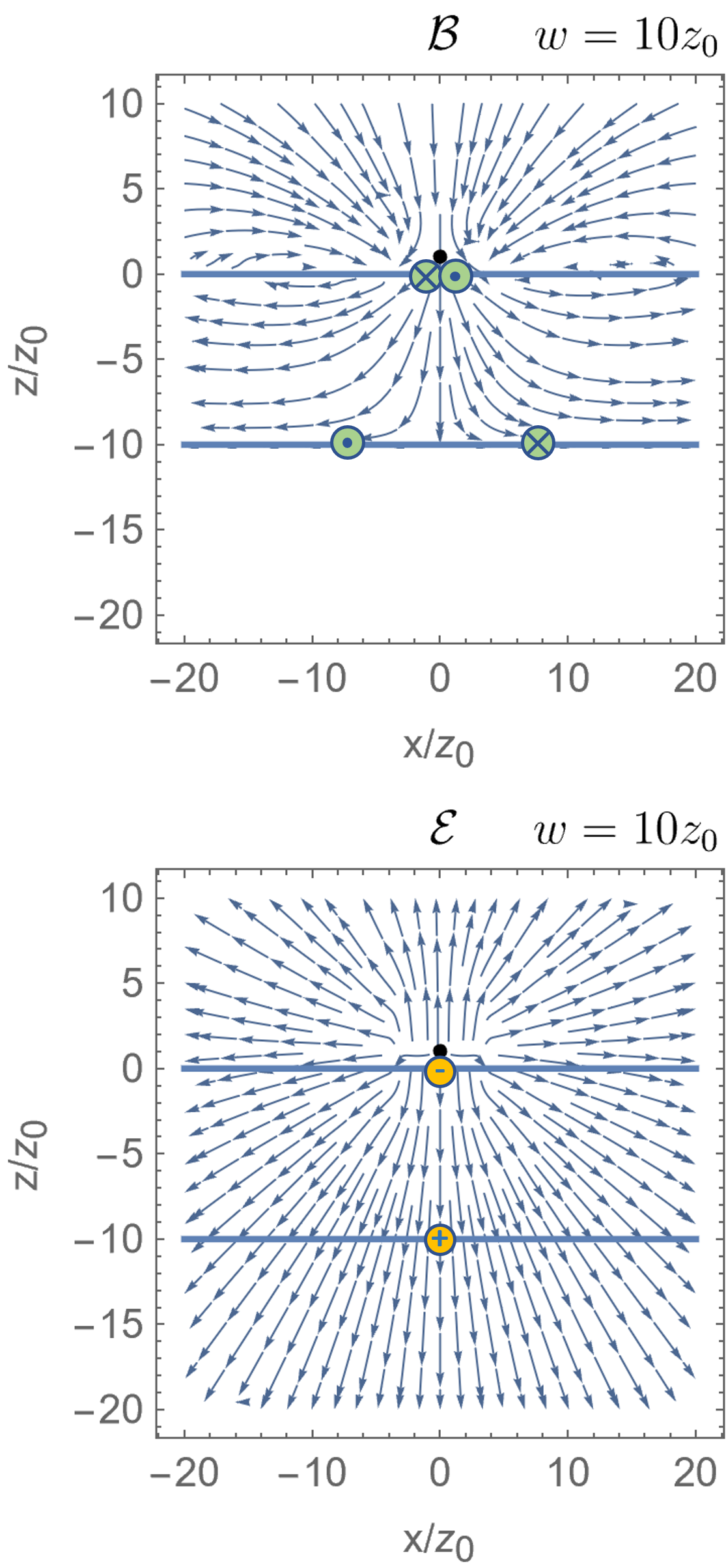}}}
\hspace{3pt}
\subfloat[Electric point charge located near a thin slab of
magnetoelectric material, i.e., $w\gtrsim z_0$.]{%
\resizebox*{4.5cm}{!}{\includegraphics{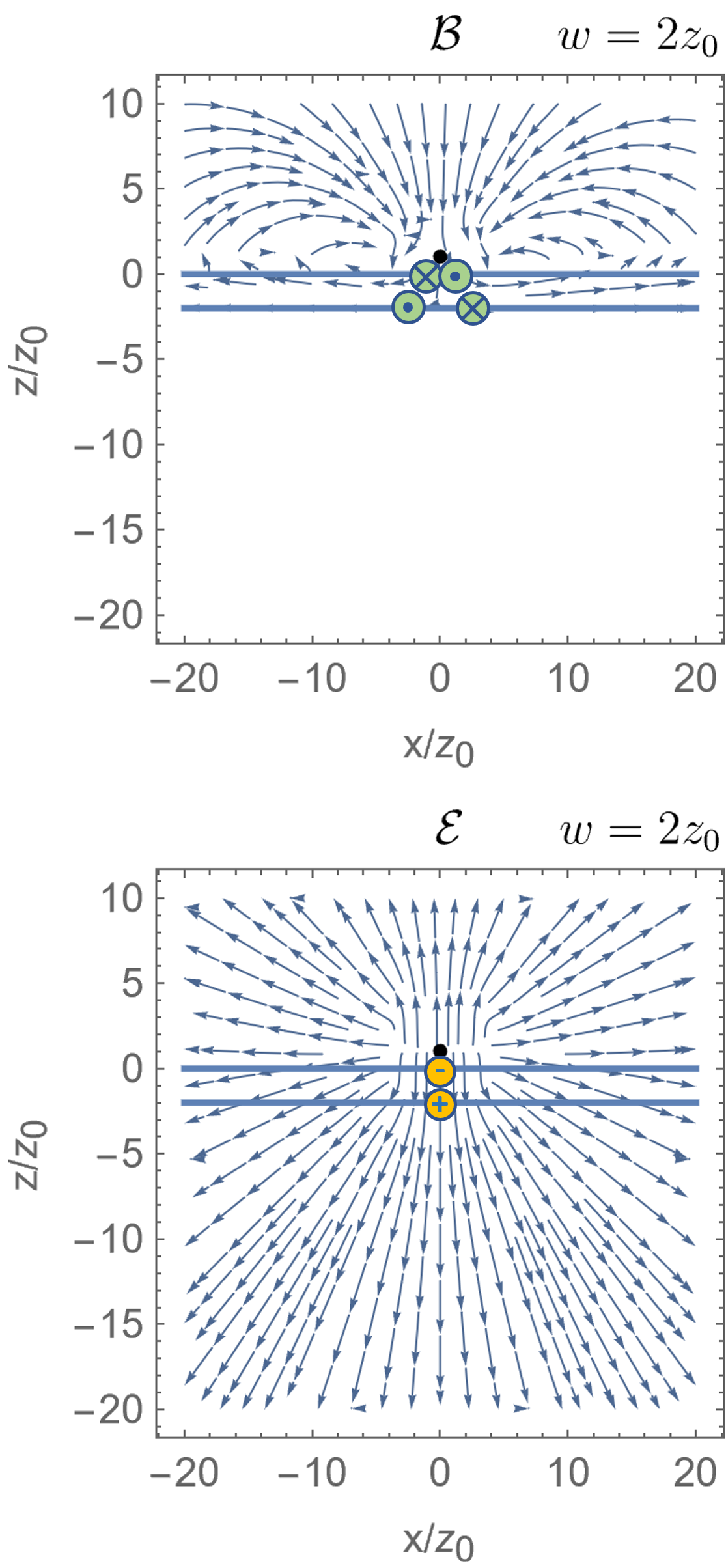}}}
\caption{\label{fig:QoutSlab}%
Field lines of the electric ($\Ec$) and magnetic ($\Bc$) fields
generated by an electric point charge located outside an
isotropic magnetoelectric medium occupying the space $0>z>-w$
and having the same dielectric constant and magnetic
permeability as the surrounding ordinary medium ($\epsilon =
\epsilon_0$, $\mu = \mu_0$). The black dot indicates the
location of the source charge, and horizontal thick blue lines
delineate interfaces between ordinary and magnetoelectric media.
Green circles are positioned where the interface-current
distribution has maxima and show the current direction. Yellow
circles are positioned where the interface-charge distribution
has maxima and show its sign. The magnetoelectric is assumed to
have $\alpha = \sqrt{\epsilon_0/\mu_0}$.}
\end{figure}

\begin{figure}
\centering
\subfloat[Electric point charge located inside a semi-infinite
magnetoelectric medium.]{%
\resizebox*{4.5cm}{!}{\includegraphics{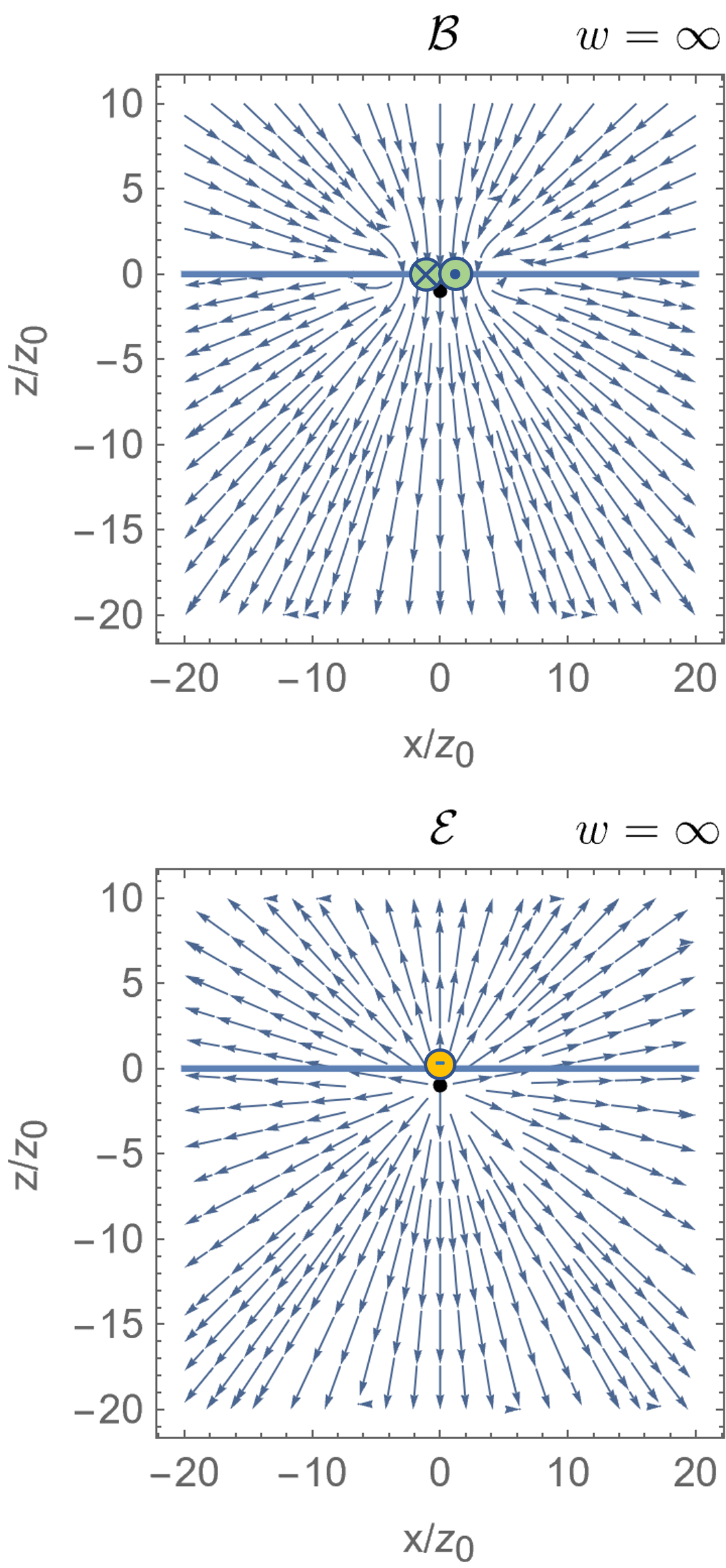}}}
\hspace{3pt}
\subfloat[Electric point charge located inside a thick slab of
magnetoelectric material, i.e., $w\gg -z_0$.]{%
\resizebox*{4.5cm}{!}{\includegraphics{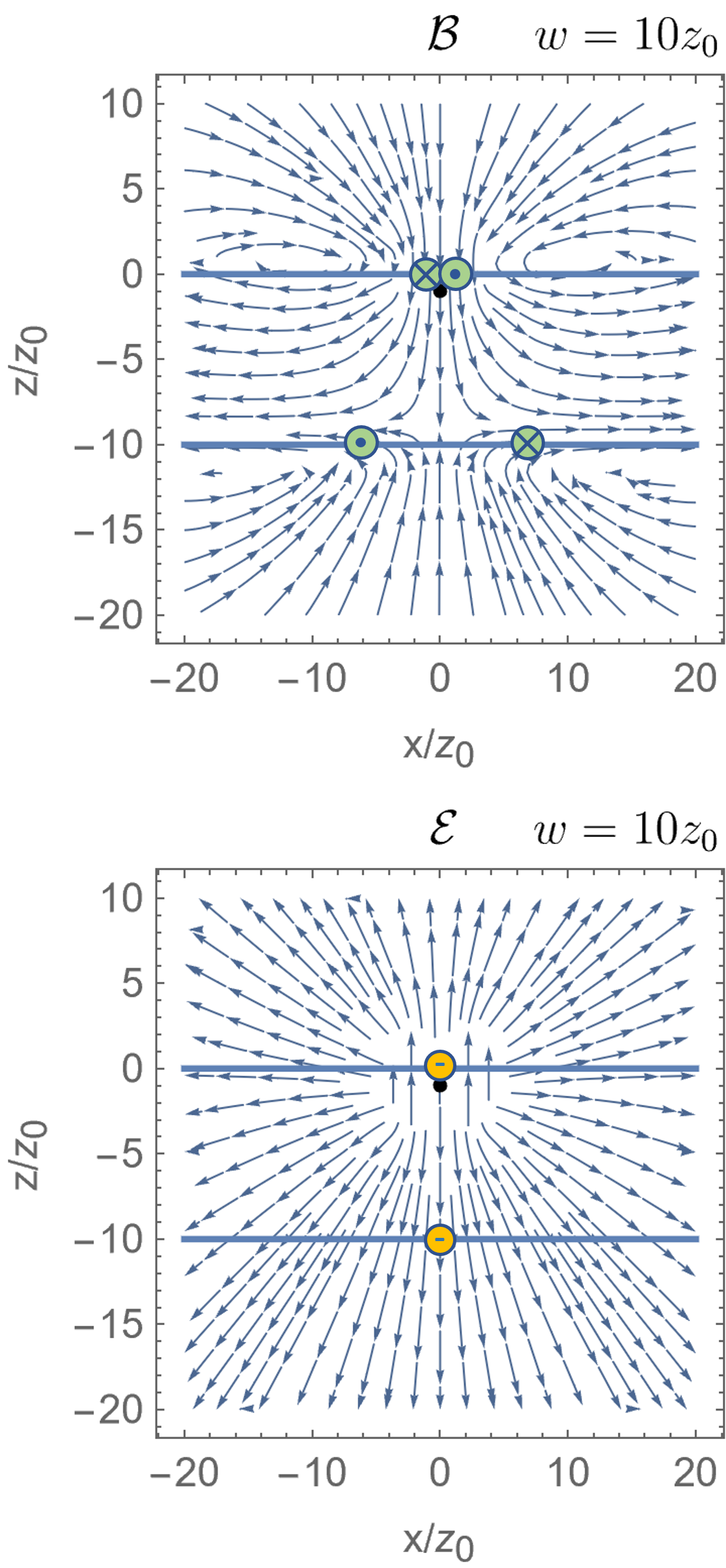}}}
\hspace{3pt}
\subfloat[Electric point charge located inside a thin slab of
magnetoelectric material, i.e., $w\gtrsim -z_0$.]{%
\resizebox*{4.5cm}{!}{\includegraphics{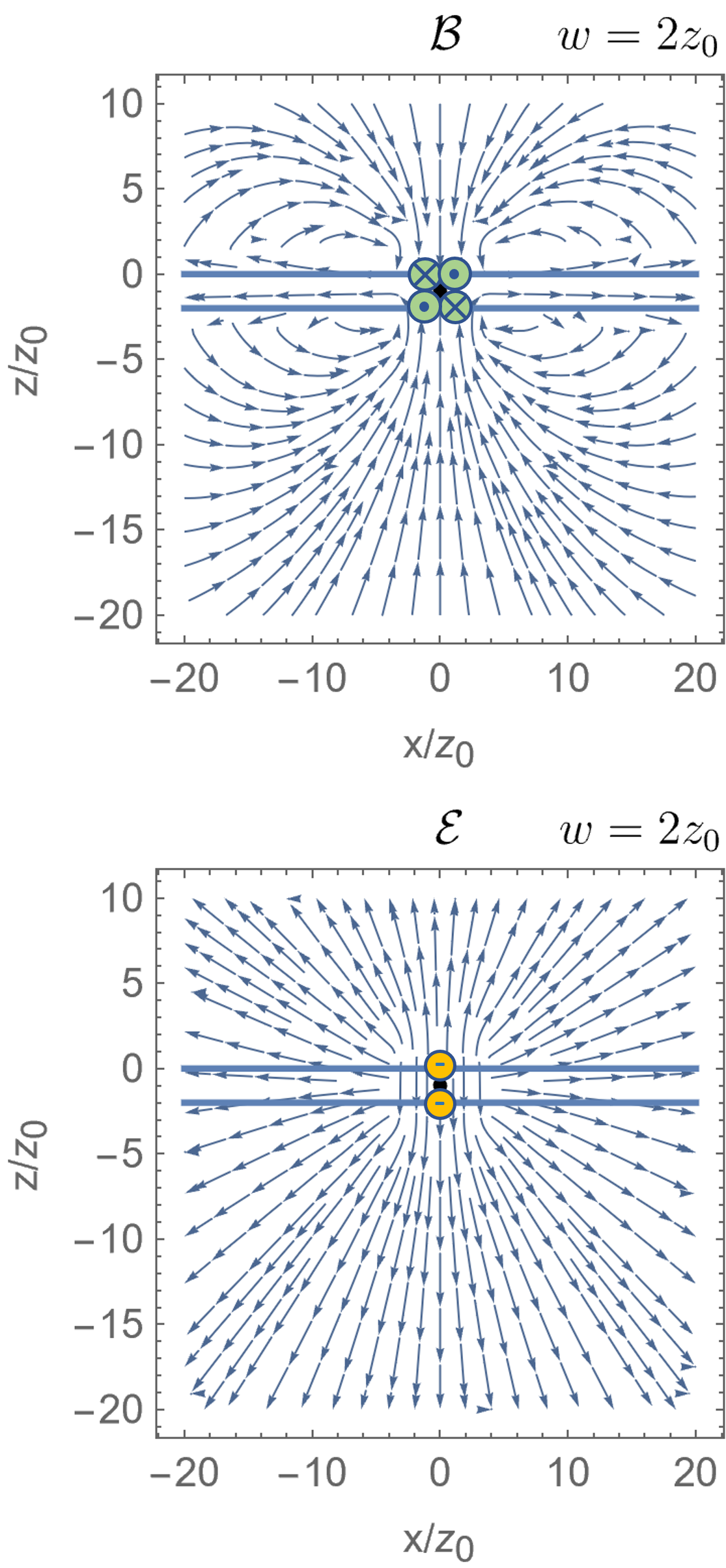}}}
\caption{\label{fig:QinSlab}%
Field lines of the electric ($\Ec$) and magnetic ($\Bc$) fields
generated by an electric point charge located inside an
isotropic magnetoelectric medium occupying the space $0>z>-w$
and having the same dielectric constant and magnetic
permeability as the surrounding ordinary medium ($\epsilon =
\epsilon_0$, $\mu = \mu_0$). The black dot indicates the
location of the source charge, and horizontal thick blue lines
delineate interfaces between ordinary and magnetoelectric media.
Green circles are positioned where the interface-current
distribution has maxima and show the current direction. Yellow
circles are positioned where the interface-charge distribution
has maxima and show its sign. The magnetoelectric is assumed to
have $\alpha = \sqrt{\epsilon_0/\mu_0}$.}
\end{figure}

\begin{figure}
\centering
\subfloat[Electric point charge located outside a semi-infinite
magnetoelectric medium.]{%
\resizebox*{4.5cm}{!}{\includegraphics{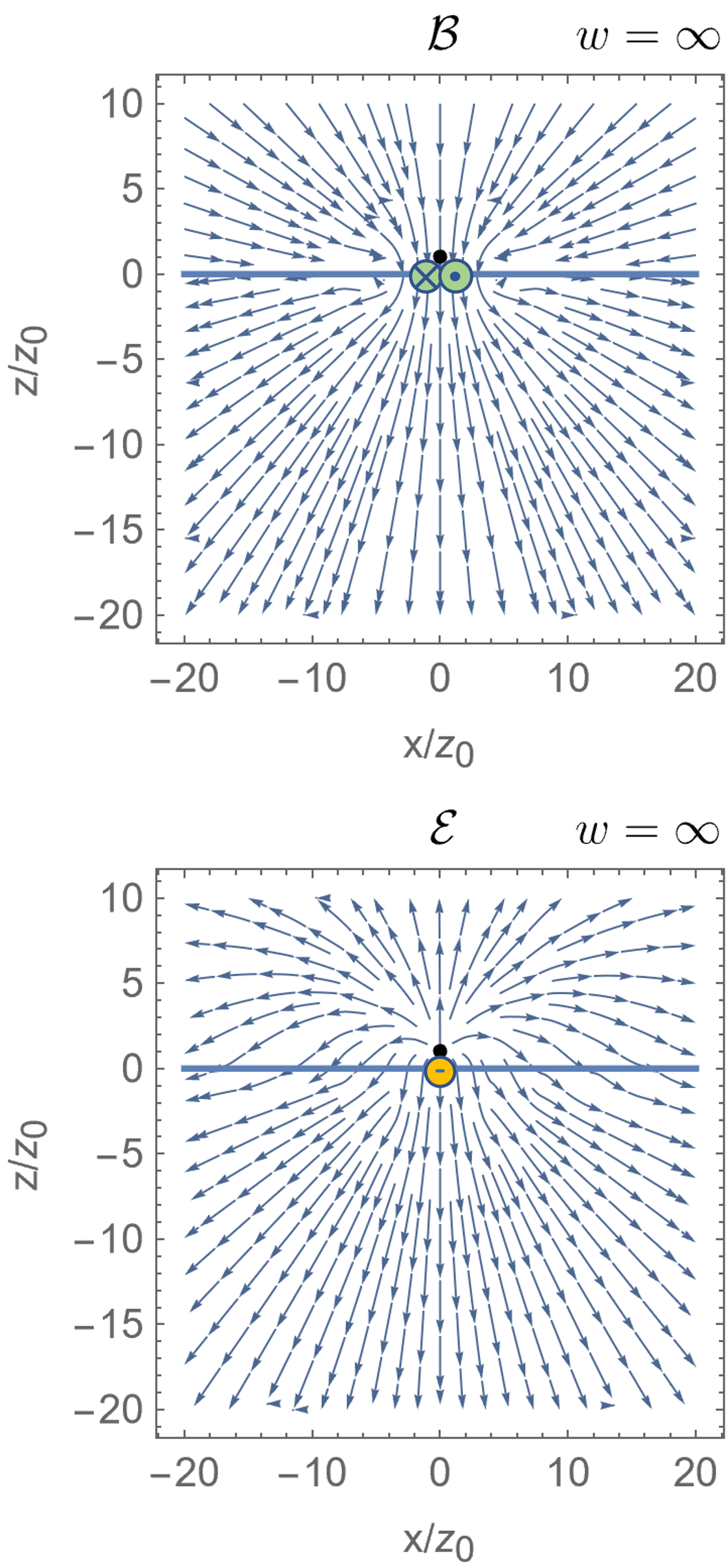}}}
\hspace{3pt}
\subfloat[Electric point charge located near a thick slab of
magnetoelectric material, i.e., $w\gg z_0$.]{%
\resizebox*{4.5cm}{!}{\includegraphics{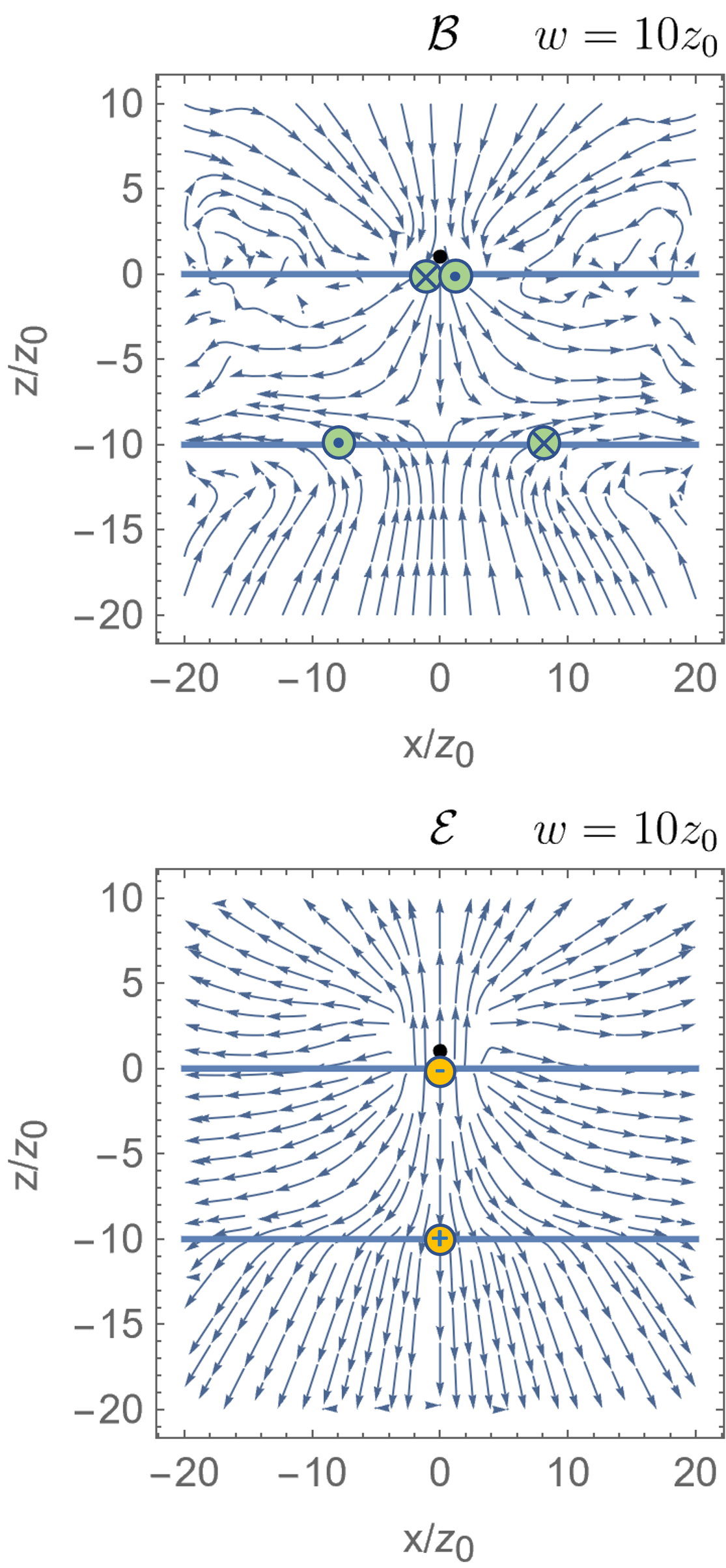}}}
\hspace{3pt}
\subfloat[Electric point charge located near a thin slab of
magnetoelectric material, i.e., $w\gtrsim z_0$.]{%
\resizebox*{4.5cm}{!}{\includegraphics{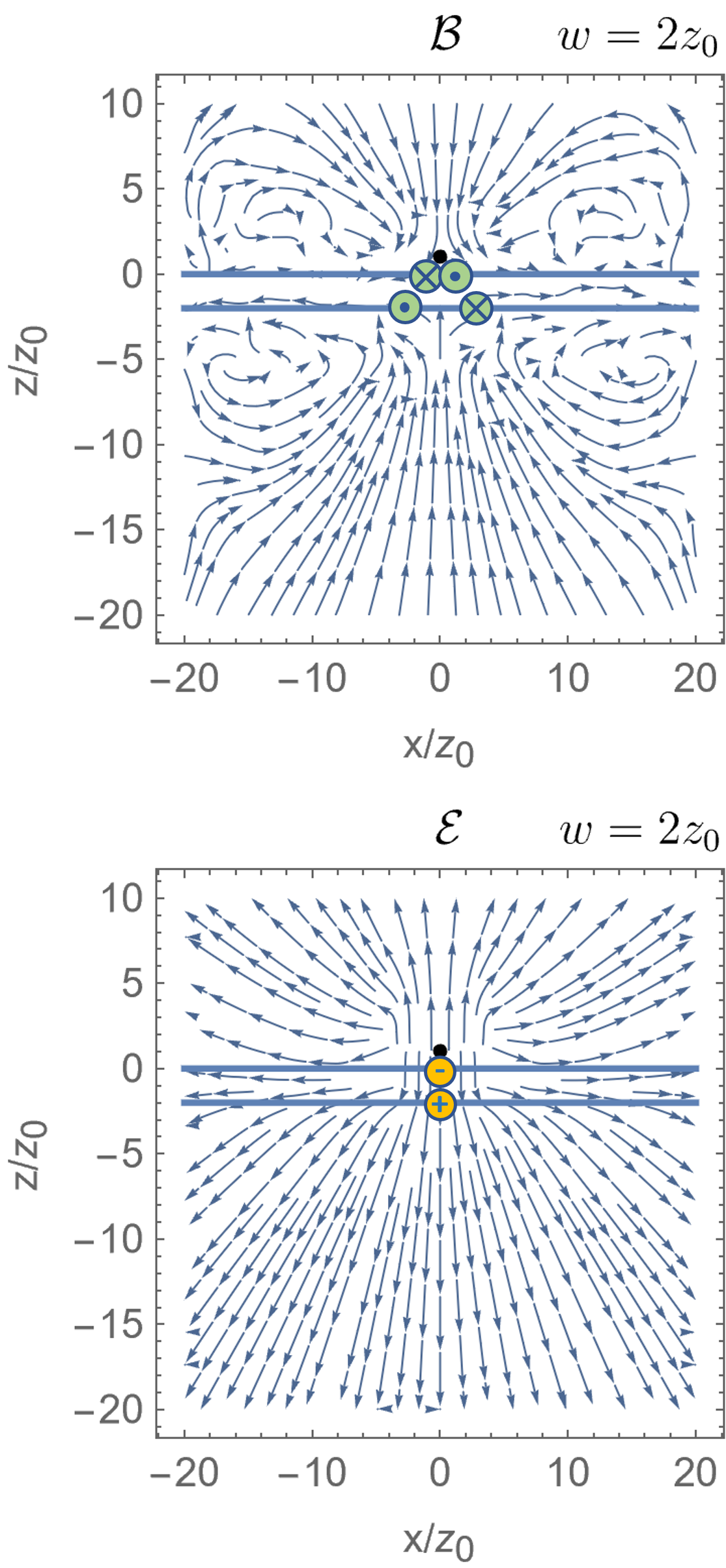}}}
\caption{\label{fig:QdoutSlabCr}%
Field lines of the electric ($\Ec$) and magnetic ($\Bc$) fields
generated by an electric point charge located outside an
isotropic magnetoelectric medium occupying the space $0>z>-w$
and having $\epsilon=10\,\epsilon_0$, $\mu=\mu_0$, and $\alpha =
3 \times10^{-4}\sqrt{\epsilon_0/\mu_0}$. The black dot indicates
the location of the source charge, and horizontal thick blue
lines delineate interfaces between ordinary and magnetoelectric
media. Green circles are positioned where the interface-current
distribution has maxima and show the current direction. Yellow
circles are positioned where the interface-charge distribution
has maxima and show its sign.}
\end{figure}

\begin{figure}
\centering
\subfloat[Charge outside a thick slab ($w=10\, z_0$) with
$\epsilon = \epsilon_0$.]{%
\resizebox*{4.4cm}{!}{\includegraphics{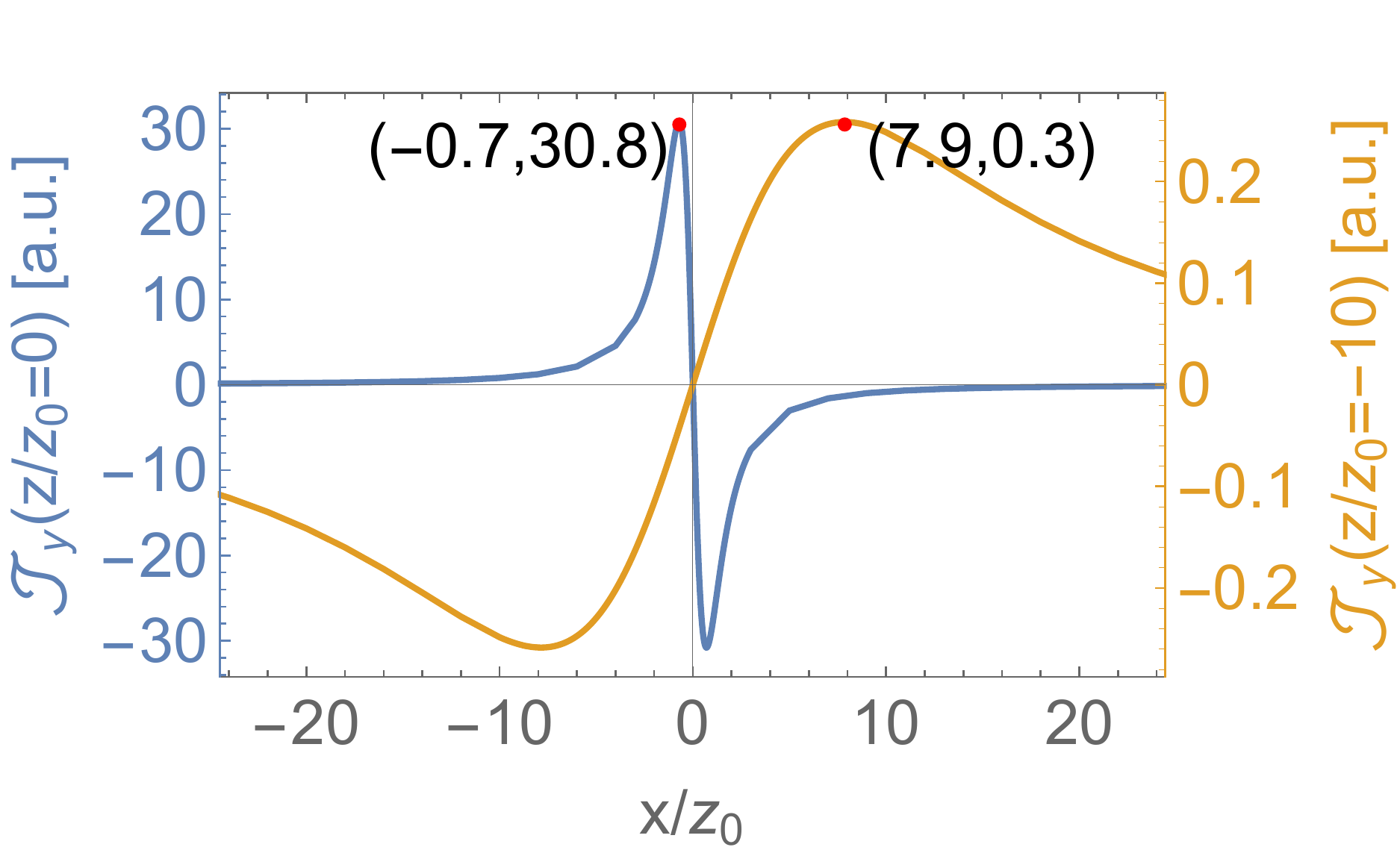}}}
\hspace{8pt}
\subfloat[Charge outside a thick slab ($w=10\, z_0$) with
$\epsilon = 10\, \epsilon_0$.]{%
\resizebox*{4.4cm}{!}{\includegraphics{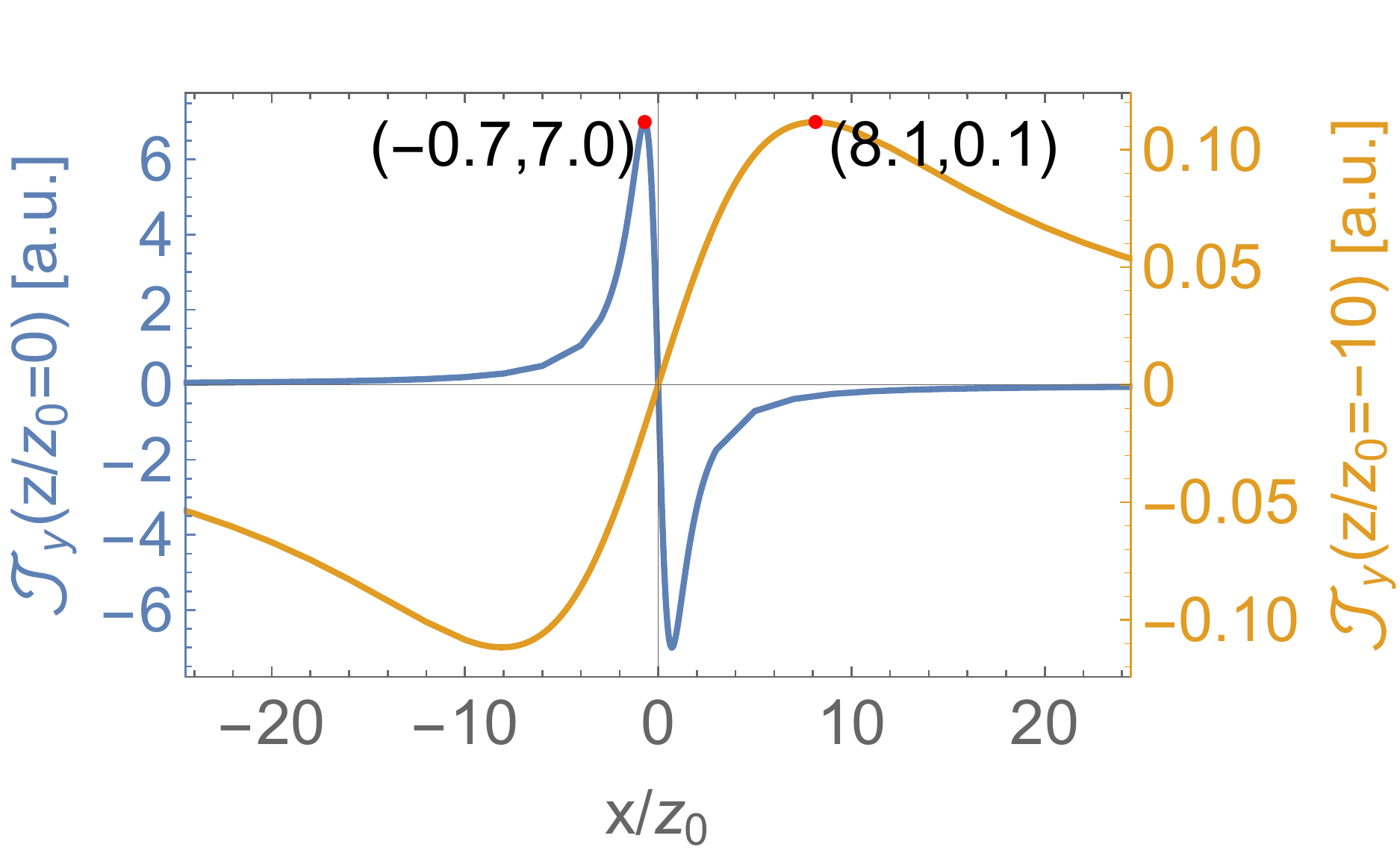}}}
\hspace{8pt}
\subfloat[Charge inside a thick slab ($w=10\, z_0$) with
$\epsilon = \epsilon_0$.]{%
\resizebox*{4.4cm}{!}{\includegraphics{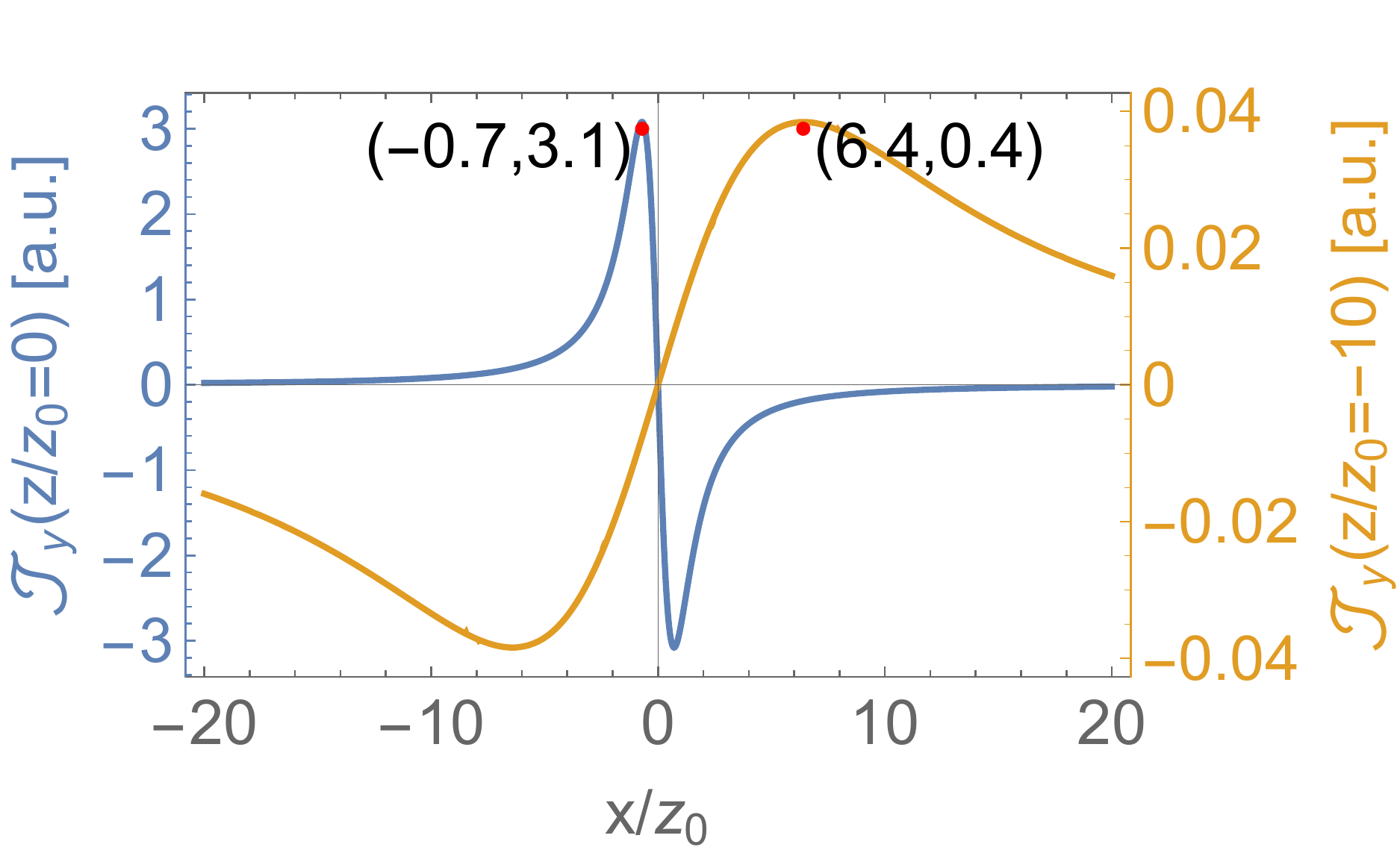}}}
\vspace{10pt}
\subfloat[Charge outside a thin slab ($w=2\, z_0$) with
$\epsilon = \epsilon_0$.]{%
\resizebox*{4.4cm}{!}{\includegraphics{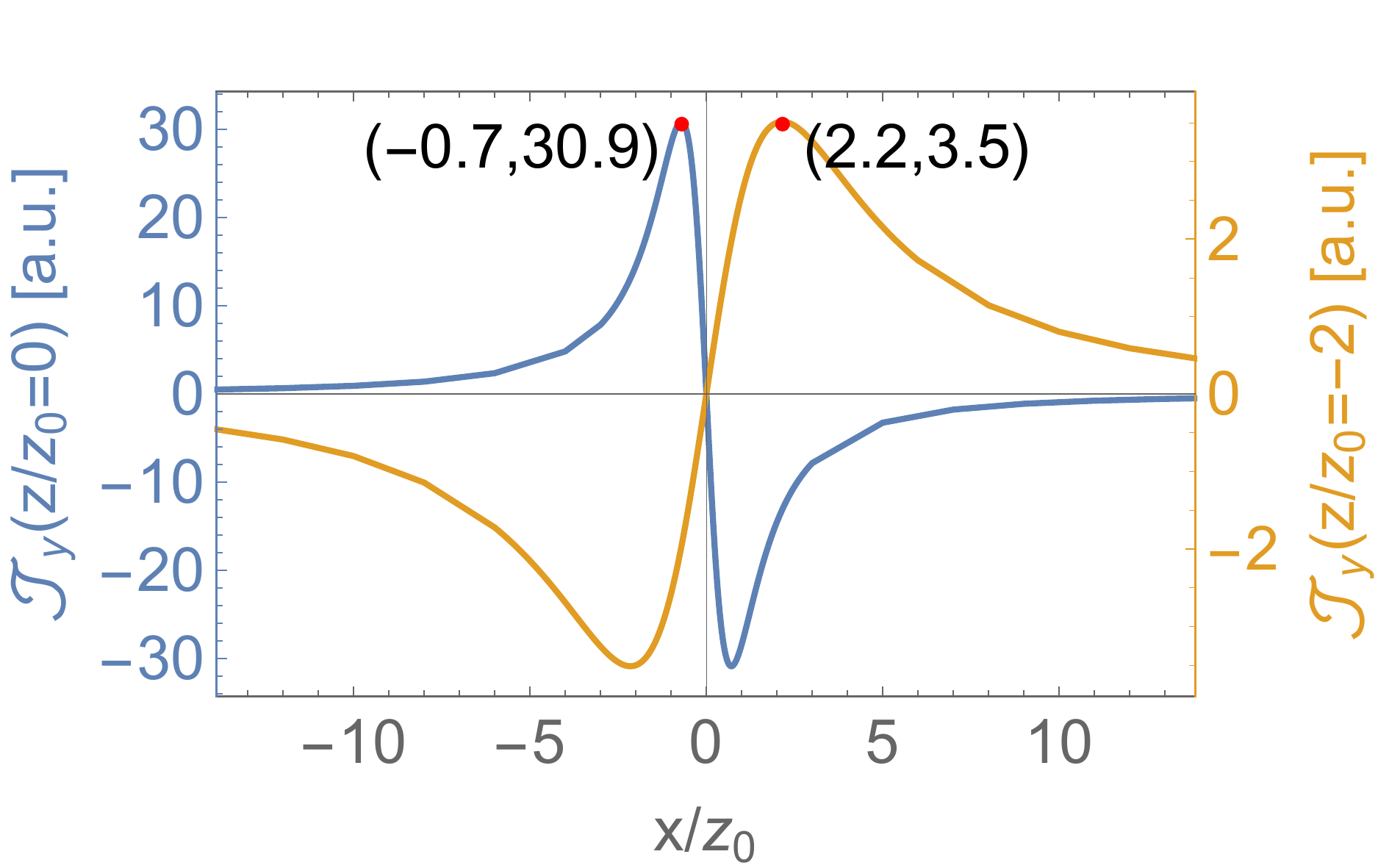}}}
\hspace{8pt}
\subfloat[Charge outside a thin slab ($w=2\, z_0$) with
$\epsilon = 10\, \epsilon_0$.]{%
\resizebox*{4.4cm}{!}{\includegraphics{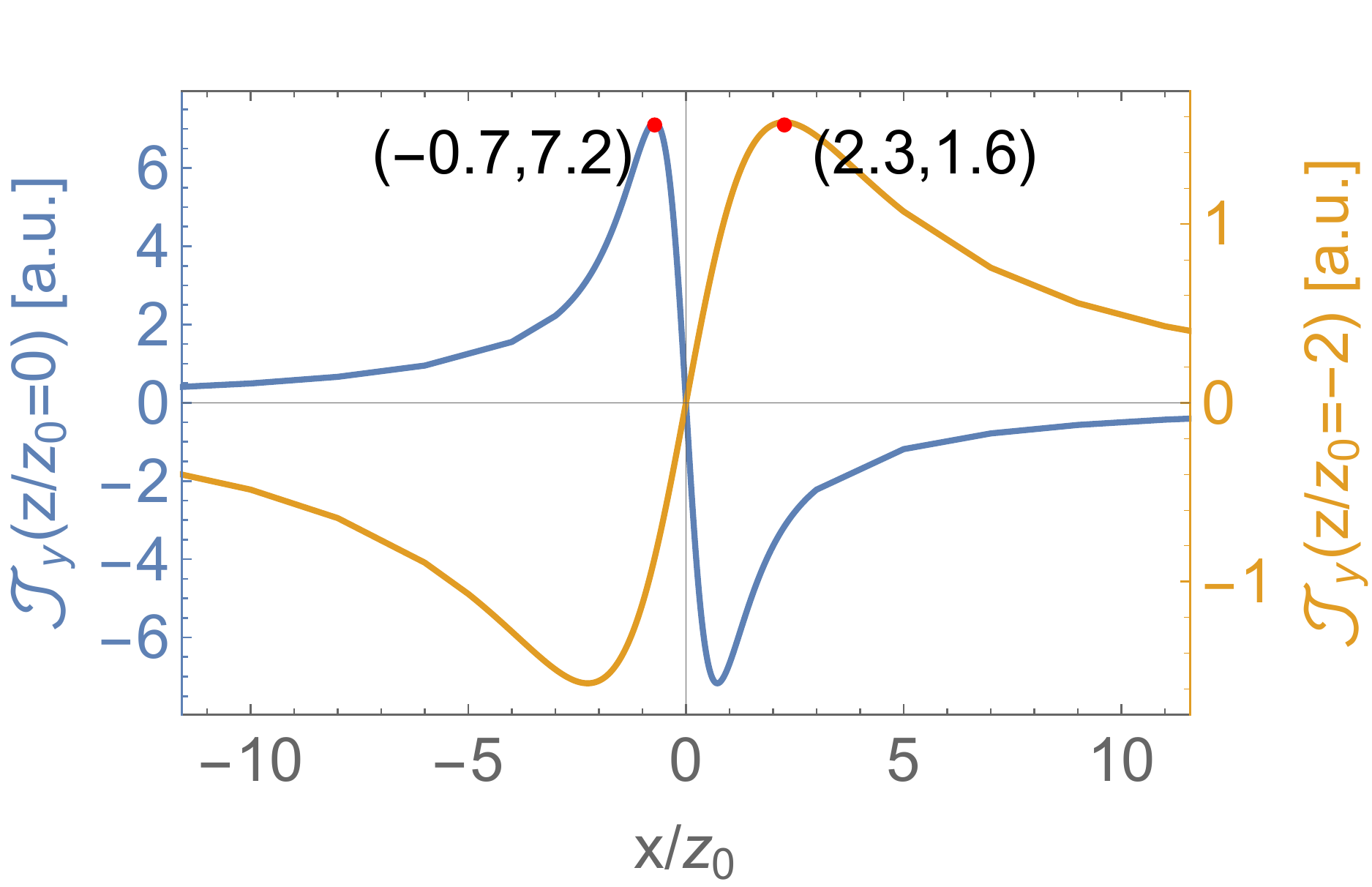}}}
\hspace{8pt}
\subfloat[Charge inside a thin slab \hspace{1em} ($w=2\, z_0$)
with $\epsilon = \epsilon_0$.]{%
\resizebox*{4.4cm}{!}{\includegraphics{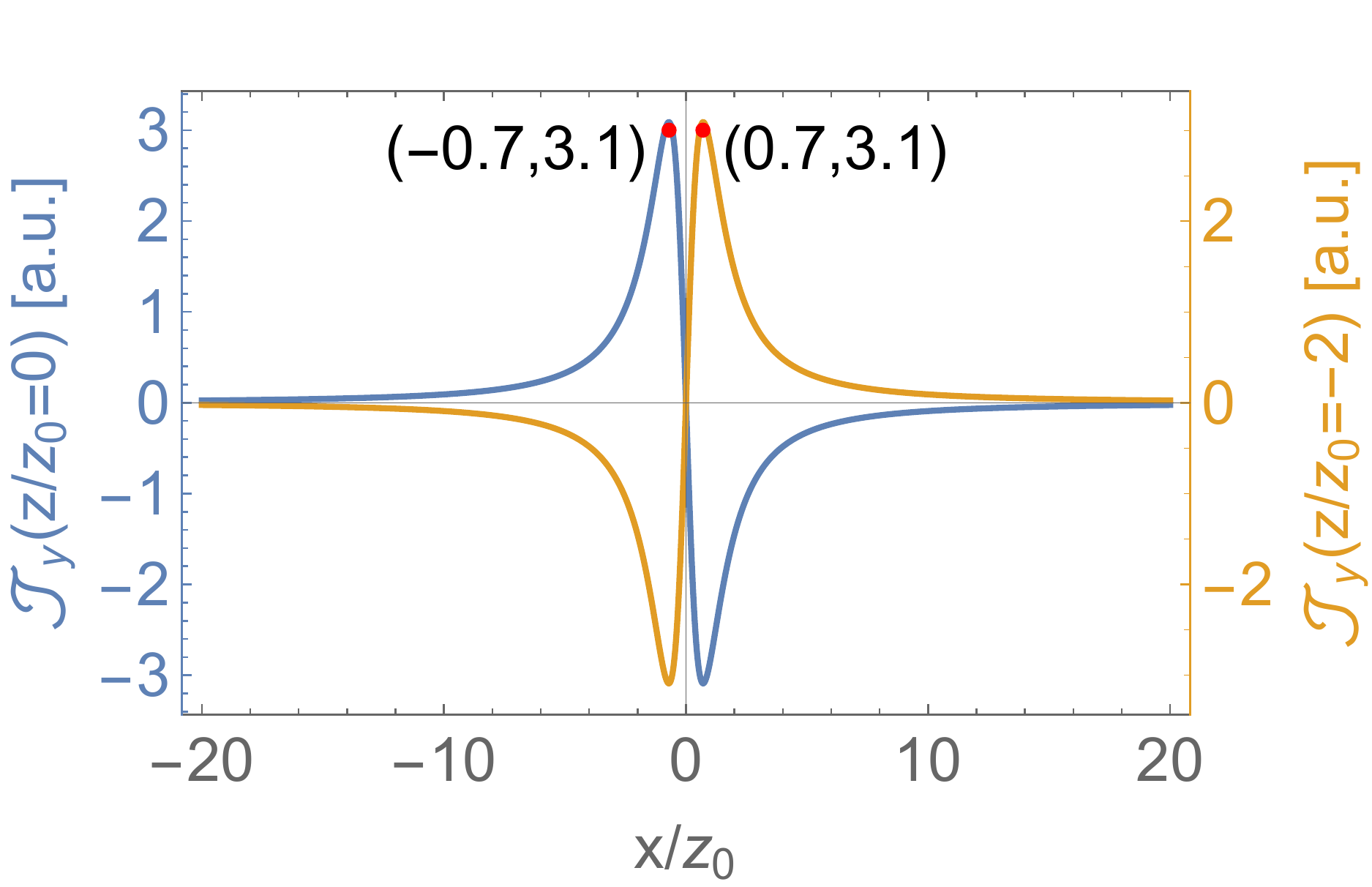}}}
\caption{\label{fig:Jout}
Magnetoelectrically induced interface currents in the slab
geometries considered in this work. The blue (yellow) curves
show the variation of the current-density component $\Jc_y$ at
the upper (lower) boundary as a function of coordinate $x/z_0$.
Pairs $(x/z_0, \Jc_y)$ of values associated with
positive-current maxima are indicated, allowing also the
negative-current maxima to be identified via symmetry as
$(-x/z_0, -\Jc_y)$. Panel~(a) [(b), (c), (d), (e), (f)] pertains
to the situations shown in Fig.~\ref{fig:QoutSlab}(b)
[Fig.~\ref{fig:QdoutSlabCr}(b), Fig.~\ref{fig:QinSlab}(b),
Fig.~\ref{fig:QoutSlab}(c), Fig.~\ref{fig:QdoutSlabCr}(c),
Fig.~\ref{fig:QinSlab}(c)].}
\end{figure}

Without loss of generality, we assume a positively charged
point source ($q_0>0$) from now on and only calculate
electromagnetic fields in the plane $\theta = 0$. As a basic
illustration of the magnetoelectric effect, we first present
results for an artificial situation where the magnetoelectric
has the same dielectric constant $\epsilon = \epsilon_0$ and
magnetic permeability $\mu = \mu_0$ as the surrounding medium;
i.e., the magnetoelectric medium only differs by having finite
$\alpha$. Fields for configurations with these parameter values
and the electric source charge located near (inside) the
magnetoelectric are shown in Fig.~\ref{fig:QoutSlab}
(Fig.~\ref{fig:QinSlab}). Results for a more realistic set of
magnetoelectric-medium parameters ($\epsilon = 10\,\epsilon_0$,
$\mu = \mu_0$, $\alpha = 3\times 10^{-4}\, \sqrt{\epsilon_0/
\mu_0}$) are shown in Fig.~\ref{fig:QdoutSlabCr}. To provide
further insight into the obtained field configurations, we
indicate the locations of maxima in the distributions of
interfacial currents and charges. Plots of the interface-current
distributions in finite-width slab geometries are shown in
Fig.~\ref{fig:Jout}.

We start by discussing the idealized situation with $\epsilon =
\epsilon_0$ and $\mu = \mu_0$, i.e., when the differentiating
characteristic of the region inside (outside) the
magnetoelectric medium is a finite (vanishing) $\alpha$. When
the entire half-space $z<0$ is occupied by the magnetoelectric,
there is only one boundary in the system, on which the surface
currents and surface charges are induced by the source electric
charge, as indicated in Figs.~\ref{fig:QoutSlab}(a) and
\ref{fig:QinSlab}(a). Since the image charges are effective
representations of the surface currents and surface charges, the
two magnetic image monopoles $g^{+}_1$ and $g^{-}_1$ located at
different sides of the boundary have the same magnitudes but
opposite signs while the two electric image charges $q^{+}_1$
and $q^{-}_1$ are identical. Therefore, when $\epsilon =
\epsilon_0$ and $\mu = \mu_0$, the magnetic field above the
boundary is non-zero and mirror-symmetric to the one below but
with opposite directions. The electric field, specifically
$\Ec_z$, is not continuous across the boundary, which is a
result of the superposition of the electric field generated by
the real source charge and the one generated by the electric
image charges, with the latter ones being symmetric with respect
to the boundary. When the width $w$ of the magnetoelectric
medium is finite and the source charge placed above it, the
magnetic fields are non-zero above and inside the slab but
vanish below, as shown in Figs.~\ref{fig:QoutSlab}(b) and
\ref{fig:QoutSlab}(c). The vanishing of the magnetic field below
the slab is due to the cancellation effect between the magnetic
fields generated by the surface currents on the upper and lower
boundaries since the directions of the induced currents are
opposite at the two boundaries. Furthermore, the magnitude of
the current density at the lower surface is weaker than the one
at the upper surface since the lower boundary is further away
from the source charge and is more spread out, as can be seen
from Fig.~\ref{fig:Jout}. The electric field has the same form
inside and below the slab, which arises due to the boundary
conditions at $z=-w$ where $\vek{\Bc}=0$. As the slab width $w$
decreases, the magnetic field more and more resembles the one
generated by a magnetic quadrupole, especially when the source
charge is located inside the magnetoelectric medium [see
Fig.~\ref{fig:QinSlab}(c)]. This is because the distributions of
the surface currents in terms of their magnitudes at the two
boundaries become more and more similar as the slab becomes
thinner (see Fig.~\ref{fig:Jout}). Therefore, we expect the
magnetic field to be the same as a quadrupole field when the
slab is effectively 2D since the upper and lower surface current
will generate the same but opposite magnetic dipole moments.

Real materials typically have a dielectric constant $\epsilon$
that is much larger than that of vacuum ($\epsilon_0$) and quite
small values of $\alpha$ measured in units of $\sqrt{\epsilon_0
/\mu_0}$. For example, in the paradigmatic magnetoelectric
Cr$_2$O$_3$, $\epsilon\approx 10\, \epsilon_0$, $\mu\approx
\mu_0$, and $\alpha=3\times10^{-4}\sqrt{\epsilon_0/
\mu_0}$~\cite{heh08a}.\endnote{Strictly speaking, the
magnetoelectric response of Cr$_2$O$_3$ is not isotropic but
uniaxial, i.e., described by a tensor $\underline{\alpha} \equiv
\mathrm{diag}(\alpha_{xx}, \alpha_{yy}, \alpha_{zz})$ with
$\alpha_{xx} = \alpha_{yy}\ne \alpha_{zz}$~\cite{fon63,lal67,
wie94}. In our model for a realistic isotropic magnetoelectric,
we use the value of $\alpha\equiv (2\, \alpha_{xx} + \alpha_{zz}
)/3$ \cite{heh08a}.} We present results for such a materials
combination in Fig.~\ref{fig:QdoutSlabCr}. In the previously
discussed case with $\epsilon=\epsilon_0$, induced charges and
currents at the interfaces arose entirely as a result of
magnetoelectricity, i.e., the piecewise-constant spatial
variation of $\alpha$ [see Eq.~(\ref{eq:axMax})]. Due to the
additional variation in $\epsilon$ between the ordinary and
magnetoelectric media, there will be an extra electric response.
Due to the magnetoelectric effect, this extra electric response
leads to an extra contribution to the interface current at the
lower boundary. As a result, the magnetic field no longer
vanishes in the region below the slab. It rather resembles the
magnetic field above the medium, especially when the slab is
thin, as illustrated in Fig.~\ref{fig:QdoutSlabCr}. 

Figure~\ref{fig:Jout} illustrates in greater detail parametric
dependencies of the dipolar interface-current distributions that
are induced magnetoelectrically by an electric source charge in
slab geometries. The presented results are calculated
brute-force by taking $\vek{\nabla}\times\vek{\Bc}$. The upper
(lower) row of plots is associated with the thick (thin) slabs,
with plots paired in the same column pertaining to the same
source-charge location and the same material combination. The
line shapes rationalize the observed quadrupolar patterns of
magnetic field lines, especially in the thin-slab limit.

\section{Conclusions}

We have reviewed the currently available variety of
theoretically proposed and experimentally realized 2D
magnetoelectric materials. Both the already well-established
2D-quantum-well systems in semiconductor heterostructures and
the more recently fabricated 2D atomic crystals are viable
platforms for the detailed exploration and possible device
application of magnetoelectricity. At present, however,
theoretical proposals vastly outnumber actual experimental
realizations. Thus a concerted effort is needed to pursue the
promising theoretical leads and overcome basic materials
challenges.

Where experimental data are available, we have quoted or
estimated the magnitudes of magnetoelectric couplings in 2D
materials. Sizable values have been demonstrated for bilayer
CrI$_3$, but most other measurements are indicative of much
weaker magnetoelectricity than is typically exhibited in
single-phase bulk magnetoelectrics. We expect that significantly
larger magnetoelectric coupling should be observed in 2D
multiferroic heterostructures, but these materials still need
to be realized.

In addition to the survey of materials, we also discussed the
implications of low dimensionality for paradigmatic
magnetoelectric responses, focusing specifically on
magnetic-monopole fields generated by electric charges near or
in magnetoelectric media. To solve the associated boundary-value
problem, we employed a generalization of the classical
image-charge method where images carry both electric and
magnetic charges, thus forming image dyons. Previously applied
only to treat a single interface between an ordinary material
and a magnetoelectric, we extended the formalism to describe
a magnetoelectric plate. Finite-size effects are shown to
importantly affect the magnetoelectric responses. These
results provide a useful guide for further experimental
exploration of magnetoelectricity in 2D materials, and they will
hopefully stimulate broader investigation of this interesting
current topic.

\section*{Acknowledgements}
Y.Y.\ thanks Jinmin Yi for helpful discussions.

\section*{Disclosure statement}

No potential conflict of interest was reported by the author(s).

\section*{Funding}

This work was supported by the Marsden Fund Council from New
Zealand government funding (Contract No. VUW1713), managed by
the Royal Society Te Ap{\= a}rangi.

%\section*{Notes}
\theendnotes

%Optional notes may be included at the beginning and/or end of a citation by the use of square brackets, e.g.\ \verb"\cite[cf.][]{Gau05}" produces `\cite[cf.][]{Gau05}', \verb"\cite[p.356]{BGC04}" produces `\cite[p.356]{BGC04}', and \verb"\cite[see][p.73-–77]{PI51}" produces `\cite[see][p.73--77]{PI51}'.

%\bibliographystyle{tfnlm}
%\bibliography{ME_effects}

\begin{thebibliography}{100}
\providecommand{\url}[1]{\normalfont{#1}}
\providecommand{\urlprefix}{Available from: }

\bibitem{ode70}
O'Dell~TH. The electrodynamics of magneto-electric media. Amsterdam:
  North-Holland; 1970.

\bibitem{lan84}
Landau~LD, Lifshitz~EM. Electrodynamics of continuous media. {Second Revised}
  ed. Oxford: Pergamon; 1984.

\bibitem{fie05}
Fiebig~M. Revival of the magnetoelectric effect. J Phys D.
  2005;\hspace{0pt}38:R123.

\bibitem{jac99}
Jackson~JD. Classical electrodynamics. 3rd ed. New York (NY): Wiley; 1999.

\bibitem{dzy59}
Dzyaloshinski{\u\i}~IE. On the magneto-electrical effect in antiferromagnets.
  Zh Eksp Teor Fiz. 1959;\hspace{0pt}37:881--882. [Sov. Phys. JETP \textbf{10},
  628 (1960].

\bibitem{sch73}
Schmid~H. On a magnetoelectric classification of materials. Int J Magn.
  1973;\hspace{0pt}4:337--361.

\bibitem{sir94}
Siratori~K. Magneto-electric effect and solid state physics. Ferroelectrics.
  1994;\hspace{0pt}161:29--41.

\bibitem{fie16}
Fiebig~M, Lottermoser~T, Meier~D, et~al. The evolution of multiferroics. Nat
  Rev Mater. 2016;\hspace{0pt}1:16046.

\bibitem{spa17}
Spaldin~NA. Multiferroics: Past, present, and future. MRS Bulletin.
  2017;\hspace{0pt}42:385--390.

\bibitem{eer06}
Eerenstein~W, Mathur~ND, Scott~JF. Multiferroic and magnetoelectric materials.
  Nature. 2006;\hspace{0pt}442(7104):759--765.

\bibitem{don15}
Dong~S, Liu~JM, Cheong~SW, et~al. Multiferroic materials and magnetoelectric
  physics: symmetry, entanglement, excitation, and topology. Adv Phys.
  2015;\hspace{0pt}64:519--626.

\bibitem{spa19}
Spaldin~NA, Ramesh~R. Advances in magnetoelectric multiferroics. Nat Mater.
  2019;\hspace{0pt}18:203--212.

\bibitem{don19}
Dong~S, Xiang~H, Dagotto~E. Magnetoelectricity in multiferroics: a theoretical
  perspective. Natl Sci Rev. 2019;\hspace{0pt}6:629--641.

\bibitem{hu17}
Hu~JM, Duan~CG, Nan~CW, et~al. Understanding and designing magnetoelectric
  heterostructures guided by computation: progresses, remaining questions, and
  perspectives. npj Comput Mater. 2017;\hspace{0pt}3:1--21.

\bibitem{chu07}
Chu~YH, Martin~LW, Holcomb~MB, et~al. Controlling magnetism with multiferroics.
  Mater Today. 2007;\hspace{0pt}10:16--23.

\bibitem{fus14}
Fusil~S, Garcia~V, Barth{\'e}l{\'e}my~A, et~al. Magnetoelectric devices for
  spintronics. Annu Rev Mater Res. 2014;\hspace{0pt}44:91--116.

\bibitem{ort15}
Ortega~N, Kumar~A, Scott~JF, et~al. Multifunctional magnetoelectric materials
  for device applications. J Phys: Condens Matter. 2015;\hspace{0pt}27:504002.

\bibitem{son17}
Song~C, Cui~B, Li~F, et~al. Recent progress in voltage control of magnetism:
  Materials, mechanisms, and performance. Prog Mater Sci.
  2017;\hspace{0pt}87:33--82.

\bibitem{mani19}
Manipatruni~S, Nikonov~DE, Lin~CC, et~al. Scalable energy-efficient
  magnetoelectric spin–orbit logic. Nature. 2019;\hspace{0pt}565:35--42.

\bibitem{sik83}
Sikivie~P. Experimental tests of the "invisible" axion. Phys Rev Lett.
  1983;\hspace{0pt}51:1415--1417.

\bibitem{wil87}
Wilczek~F. Two applications of axion electrodynamics. Phys Rev Lett.
  1987;\hspace{0pt}58:1799--1802.

\bibitem{heh08a}
Hehl~FW, Obukhov~YN, Rivera~JP, et~al. Relativistic nature of a magnetoelectric
  modulus of {Cr}$_{2}${O}$_{3}$ crystals: A four-dimensional pseudoscalar and
  its measurement. Phys Rev A. 2008;\hspace{0pt}77:022106.

\bibitem{qi08}
Qi~XL, Hughes~TL, Zhang~SC. Topological field theory of time-reversal invariant
  insulators. Phys Rev B. 2008;\hspace{0pt}78:195424.

\bibitem{ess09}
Essin~AM, Moore~JE, Vanderbilt~D. Magnetoelectric polarizability and axion
  electrodynamics in crystalline insulators. Phys Rev Lett.
  2009;\hspace{0pt}102:146805.

\bibitem{arm19}
Armitage~NP, Wu~L. On the matter of topological insulators as magnetoelectrics.
  SciPost Phys. 2019;\hspace{0pt}6:46.

\bibitem{nen20}
Nenno~DM, Garcia~CAC, Gooth~J, et~al. Axion physics in condensed-matter
  systems. Nat Rev Phys. 2020;\hspace{0pt}2:682--696.

\bibitem{sek21}
Sekine~A, Nomura~K. Axion electrodynamics in topological materials. J Appl
  Phys. 2021;\hspace{0pt}129:141101.

\bibitem{ma15}
Ma~J, Pesin~DA. Chiral magnetic effect and natural optical activity in metals
  with or without {W}eyl points. Phys Rev B. 2015;\hspace{0pt}92:235205.

\bibitem{den21}
Deng~K, Van~Dyke~JS, Minic~D, et~al. Exploring self-consistency of the
  equations of axion electrodynamics in {W}eyl semimetals. Phys Rev B.
  2021;\hspace{0pt}104:075202.

\bibitem{qi09}
Qi~XL, Li~R, Zang~J, et~al. Inducing a magnetic monopole with topological
  surface states. Science. 2009;\hspace{0pt}323:1184--1187.

\bibitem{fec14}
Fechner~M, Spaldin~NA, Dzyaloshinskii~IE. Magnetic field generated by a charge
  in a uniaxial magnetoelectric material. Phys Rev B.
  2014;\hspace{0pt}89:184415.

\bibitem{mei19}
Meier~QN, Fechner~M, Nozaki~T, et~al. Search for the magnetic monopole at a
  magnetoelectric surface. Phys Rev X. 2019;\hspace{0pt}9:011011.

\bibitem{mar16}
Mart\'{\i}n-Ruiz~A, Cambiaso~M, Urrutia~LF. Electro- and magnetostatics of
  topological insulators as modeled by planar, spherical, and cylindrical
  $\ensuremath{\theta}$ boundaries: {G}reen's function approach. Phys Rev D.
  2016;\hspace{0pt}93:045022.

\bibitem{mar21}
Mart{\'i}n-Ruiz~A, Cambiaso~M, Urrutia~LF. Axion electrodynamics in
  magnetoelectric media. In: Kamenetskii~E, editor. Chirality, magnetism and
  magnetoelectricity. (Topics in Applied Physics; Vol. 138). Cham: Springer;
  2021. p. 459--492.

\bibitem{oue19}
Ouellet~J, Bogorad~Z. Solutions to axion electrodynamics in various geometries.
  Phys Rev D. 2019;\hspace{0pt}99:055010.

\bibitem{och12}
Ochiai~T. Theory of light scattering in axion electrodynamics. J Phys Soc Jpn.
  2012;\hspace{0pt}81:094401.

\bibitem{kho14}
Khomskii~DI. Magnetic monopoles and unusual dynamics of magnetoelectrics. Nat
  Commun. 2014;\hspace{0pt}5:4793.

\bibitem{kho21}
Khomskii~DI. Multiferroics and beyond: {Electric} properties of different
  magnetic textures. J Exp Theor Phys. 2021;\hspace{0pt}132:482--492.

\bibitem{uri20}
Uri~A, Kim~Y, Bagani~K, et~al. Nanoscale imaging of equilibrium quantum {Hall}
  edge currents and of the magnetic monopole response in graphene. Nat Phys.
  2020;\hspace{0pt}16:164--170.

\bibitem{kam20}
Kamenetskii~EO. Electrodynamics of magnetoelectric media and magnetoelectric
  fields. Ann Phys (Berlin). 2020;\hspace{0pt}532:1900423.

\bibitem{vel11}
Velev~JP, Jaswal~SS, Tsymbal~EY. Multi-ferroic and magnetoelectric materials
  and interfaces. Phil Trans R Soc A. 2011;\hspace{0pt}369:3069--3097.

\bibitem{hu15}
Hu~JM, Nan~T, Sun~NX, et~al. Multiferroic magnetoelectric nanostructures for
  novel device applications. MRS Bull. 2015;\hspace{0pt}40:728--735.

\bibitem{nov05}
Novoselov~KS, Jiang~D, Schedin~F, et~al. Two-dimensional atomic crystals. Proc
  Natl Acad Sci USA. 2005;\hspace{0pt}102:10451--10453.

\bibitem{nov16}
Novoselov~KS, Mishchenko~A, Carvalho~A, et~al. {2D} materials and van der
  {Waals} heterostructures. Science. 2016;\hspace{0pt}353:aac9439.

\bibitem{liu16}
Liu~Y, Weiss~NO, Duan~X, et~al. Van der {Waals} heterostructures and devices.
  Nat Rev Mater. 2016;\hspace{0pt}1:1--17.

\bibitem{sie21}
Sierra~JF, Fabian~J, Kawakami~RK, et~al. Van der {Waals} heterostructures for
  spintronics and opto-spintronics. Nat Nanotechnol.
  2021;\hspace{0pt}16:856--868.

\bibitem{bau84}
Bauer~G, Kuchar~F, Heinrich~H, editors. Two-dimensional systems,
  heterostructures, and superlattices. (Springer Series in Solid-State
  Sciences; Vol.~53). Berlin: Springer; 1984.

\bibitem{dav98}
Davies~JH. The physics of low-dimensional semiconductors. Cambridge, UK:
  Cambridge U Press; 1998.

\bibitem{him99}
Himpsel~FJ. Magnetic quantum wells. J Phys: Condens Matter.
  1999;\hspace{0pt}11:9483--9494.

\bibitem{die14}
Dietl~T, Ohno~H. Dilute ferromagnetic semiconductors: Physics and spintronic
  structures. Rev Mod Phys. 2014;\hspace{0pt}86:187--251.

\bibitem{gib19}
Gibertini~M, Koperski~M, Morpurgo~AF, et~al. Magnetic {2D} materials and
  heterostructures. Nat Nanotechnol. 2019;\hspace{0pt}14:408--419.

\bibitem{gon19a}
Gong~C, Zhang~X. Two-dimensional magnetic crystals and emergent heterostructure
  devices. Science. 2019;\hspace{0pt}363:eaav4450.

\bibitem{mak19}
Mak~KF, Shan~J, Ralph~DC. Probing and controlling magnetic states in {2D}
  layered magnetic materials. Nat Rev Phys. 2019;\hspace{0pt}1:646--661.

\bibitem{wei20}
Wei~S, Liao~X, Wang~C, et~al. Emerging intrinsic magnetism in two-dimensional
  materials: theory and applications. 2D Mater. 2020;\hspace{0pt}8:012005.

\bibitem{lu19}
Lu~C, Wu~M, Lin~L, et~al. {Single-phase multiferroics: new materials,
  phenomena, and physics}. Natl Sci Rev. 2019;\hspace{0pt}6:653--668.

\bibitem{tan19}
Tang~X, Kou~L. Two-dimensional ferroics and multiferroics: Platforms for new
  physics and applications. J Phys Chem Lett. 2019;\hspace{0pt}10:6634--6649.

\bibitem{zho20}
Zhong~T, Li~X, Wu~M, et~al. {Room-temperature multiferroicity and diversified
  magnetoelectric couplings in {2D} materials}. Natl Sci Rev.
  2020;\hspace{0pt}7:373--380.

\bibitem{gao21a}
Gao~Y, Gao~M, Lu~Y. Two-dimensional multiferroics. Nanoscale.
  2021;\hspace{0pt}13:19324--19340.

\bibitem{pou21}
Pournaghavi~N, Pertsova~A, MacDonald~AH, et~al. Nonlocal sidewall response and
  deviation from exact quantization of the topological magnetoelectric effect
  in axion-insulator thin films. Phys Rev B. 2021;\hspace{0pt}104:L201102.

\bibitem{asc68}
Ascher~E. Higher-order magneto-electric effects. Phil Mag.
  1968;\hspace{0pt}17:149--157.

\bibitem{gri94}
Grimmer~H. The forms of tensors describing magnetic, electric and toroidal
  properties. Ferroelectrics. 1994;\hspace{0pt}161:181--189.

\bibitem{nye57}
Nye~JF. Physical properties of crystals. Oxford: Oxford University Press; 1957.

\bibitem{gan19}
Ganichev~SD, Trushin~M, Schliemann~J. Spin polarization by current. In:
  Tsymbal~EY, \v{Z}uti\'{c}~I, editors. Spintronics handbook: Spin transport
  and magnetism. 2nd ed.; Vol.~2; Chapter~7. Boca Raton: CRC Press; 2019. p.
  317--338.

\bibitem{man19}
Manchon~A, \v{Z}elezn\'y~J, Miron~IM, et~al. Current-induced spin-orbit torques
  in ferromagnetic and antiferromagnetic systems. Rev Mod Phys.
  2019;\hspace{0pt}91:035004.

\bibitem{ivc78}
Ivchenko~EL, Pikus~GE. New photogalvanic effect in gyrotropic crystals. Pis'ma
  Zh Eksp Teo Fiz. 1978;\hspace{0pt}27(11):640--643. [JETP Lett. \textbf{27},
  604 (1978)].

\bibitem{bel78}
Belinicher~VI. Space-oscillating photocurrent in crystals without symmetry
  center. Phys Lett A. 1978 May;\hspace{0pt}66(3):213--214.

\bibitem{aro91}
Aronov~AG, Lyanda-Geller~YB, Pikus~GE. Spin polarization of electrons by an
  electric current. Sov Phys-JETP. 1991;\hspace{0pt}73(3):537--541.

\bibitem{ede90}
Edelstein~VM. Spin polarization of conduction electrons induced by electron
  current in two-dimensional asymmetric electron systems. Solid State Commun.
  1990;\hspace{0pt}73(3):233--235.

\bibitem{lev85}
Levitov~LS, Nazarov~YV, {\'E}liashberg~GM. Magnetoelectric effects in
  conductors with mirror isomer symmetry. Zh Eksp Teor Fiz.
  1985;\hspace{0pt}88:229--236. [Sov. Phys. JETP \textbf{61}, 133 (1985)].

\bibitem{he20}
He~WY, Goldhaber-Gordon~D, Law~KT. Giant orbital magnetoelectric effect and
  current-induced magnetization switching in twisted bilayer graphene. Nat
  Commun. 2020;\hspace{0pt}11:1650.

\bibitem{joh21}
Johansson~A, G\"obel~B, Henk~J, et~al. Spin and orbital {E}delstein effects in
  a two-dimensional electron gas: Theory and application to
  {${\mathrm{SrTiO}}_{3}$} interfaces. Phys Rev Research.
  2021;\hspace{0pt}3:013275.

\bibitem{joh19}
Johansen~O, Risingg\aa{}rd~V, Sudb{\o}~A, et~al. Current control of magnetism
  in two-dimensional {${\mathrm{Fe}}_{3}{\mathrm{GeTe}}_{2}$}. Phys Rev Lett.
  2019;\hspace{0pt}122:217203.

\bibitem{xue21}
Xue~F, Haney~PM. Intrinsic staggered spin-orbit torque for the electrical
  control of antiferromagnets: Application to {${\mathrm{CrI}}_{3}$}. Phys Rev
  B. 2021;\hspace{0pt}104:224414.

\bibitem{lee17}
Lee~J, Wang~Z, Xie~H, et~al. Valley magnetoelectricity in single-layer {MoS2}.
  Nat Mater. 2017;\hspace{0pt}16:887--891.

\bibitem{xu14}
Xu~X, Yao~W, Xiao~D, et~al. Spin and pseudospins in layered transition metal
  dichalcogenides. Nat Phys. 2014;\hspace{0pt}10:343--350.

\bibitem{rad62}
Rado~GT. Statistical theory of magnetoelectric effects in antiferromagnetics.
  Phys Rev. 1962;\hspace{0pt}128:2546--2556.

\bibitem{tin64}
Tinkham~M. Group theory and quantum mechanics. New York: McGraw-Hill; 1964.

\bibitem{rad84}
Rado~GT, Ferrari~JM, Maisch~WG. Magnetoelectric susceptibility and magnetic
  symmetry of magnetoelectrically annealed {TbPO}$_{4}$. Phys Rev B.
  1984;\hspace{0pt}29:4041--4048.

\bibitem{riv09}
Rivera~JP. A short review of the magnetoelectric effect and related
  experimental techniques on single phase (multi-) ferroics. Eur Phys J B.
  2009;\hspace{0pt}71(3):299.

\bibitem{res10a}
Ressouche~E, Loire~M, Simonet~V, et~al. Magnetoelectric {MnPS}$_3$ as a
  candidate for ferrotoroidicity. Phys Rev B. 2010;\hspace{0pt}82:100408.

\bibitem{kur13}
Kurumaji~T, Seki~S, Ishiwata~S, et~al. Magnetoelectric responses induced by
  domain rearrangement and spin structural change in triangular-lattice
  helimagnets {NiI${}_{2}$} and {CoI${}_{2}$}. Phys Rev B.
  2013;\hspace{0pt}87:014429.

\bibitem{chu20}
Chu~H, Roh~CJ, Island~JO, et~al. Linear magnetoelectric phase in ultrathin
  {MnPS}$_3$ probed by optical second harmonic generation. Phys Rev Lett.
  2020;\hspace{0pt}124:027601.

\bibitem{lon20}
Long~G, Henck~H, Gibertini~M, et~al. Persistence of magnetism in atomically
  thin {MnPS$_3$} crystals. Nano Lett. 2020;\hspace{0pt}20:2452--2459.

\bibitem{ju21}
Ju~H, Lee~Y, Kim~KT, et~al. Possible persistence of multiferroic order down to
  bilayer limit of van der {Waals} material {NiI$_2$}. Nano Lett.
  2021;\hspace{0pt}21:5126--5132.

\bibitem{ni21}
Ni~Z, Haglund~AV, Wang~H, et~al. Imaging the {N\'eel} vector switching in the
  monolayer antiferromagnet {MnPSe$_3$} with strain-controlled {Ising} order.
  Nat Nanotechnol. 2021;\hspace{0pt}16.

\bibitem{chi16}
Chittari~BL, Park~Y, Lee~D, et~al. Electronic and magnetic properties of
  single-layer {$M\mathrm{P}{X}_{3}$} metal phosphorous trichalcogenides. Phys
  Rev B. 2016;\hspace{0pt}94:184428.

\bibitem{mcg20}
McGuire~MA. Cleavable magnetic materials from van der {Waals} layered
  transition metal halides and chalcogenides. J Appl Phys.
  2020;\hspace{0pt}128:110901.

\bibitem{jia18}
Jiang~S, Shan~J, Mak~KF. Electric-field switching of two-dimensional van der
  {Waals} magnets. Nat Mater. 2018;\hspace{0pt}17:406--410.

\bibitem{hua18}
Huang~B, Clark~G, Klein~DR, et~al. Electrical control of {2D} magnetism in
  bilayer {CrI$_3$}. Nat Nanotechnol. 2018;\hspace{0pt}13:544--548.

\bibitem{jia18a}
Jiang~S, Li~L, Wang~Z, et~al. Controlling magnetism in {2D} {CrI${}_3$} by
  electrostatic doping. Nat Nanotechnol. 2018;\hspace{0pt}13:549--553.

\bibitem{siv18}
Sivadas~N, Okamoto~S, Xu~X, et~al. Stacking-dependent magnetism in bilayer
  {CrI}$_3$. Nano Lett. 2018;\hspace{0pt}18:7658--7664.

\bibitem{lei21}
Lei~C, Chittari~BL, Nomura~K, et~al. Magnetoelectric response of
  antiferromagnetic {CrI}$_3$ bilayers. Nano Lett.
  2021;\hspace{0pt}21:1948--1954.

\bibitem{hil00}
Hill~NA. Why are there so few magnetic ferroelectrics? J Phys Chem B.
  2000;\hspace{0pt}104:6694--6709.

\bibitem{spa20}
Spaldin~NA. Multiferroics beyond electric-field control of magnetism. Proc R
  Soc A. 2020;\hspace{0pt}476:20190542.

\bibitem{luo17}
Luo~W, Xu~K, Xiang~H. Two-dimensional hyperferroelectric metals: A different
  route to ferromagnetic-ferroelectric multiferroics. Phys Rev B.
  2017;\hspace{0pt}96:235415.

\bibitem{li17}
Li~L, Wu~M. Binary compound bilayer and multilayer with vertical polarizations:
  Two-dimensional ferroelectrics, multiferroics, and nanogenerators. ACS Nano.
  2017;\hspace{0pt}11:6382--6388.

\bibitem{hua18a}
Huang~C, Du~Y, Wu~H, et~al. Prediction of intrinsic ferromagnetic
  ferroelectricity in a transition-metal halide monolayer. Phys Rev Lett.
  2018;\hspace{0pt}120:147601.

\bibitem{qi18}
Qi~J, Wang~H, Chen~X, et~al. Two-dimensional multiferroic semiconductors with
  coexisting ferroelectricity and ferromagnetism. Appl Phys Lett.
  2018;\hspace{0pt}113:043102.

\bibitem{lai19}
Lai~Y, Song~Z, Wan~Y, et~al. Two-dimensional ferromagnetism and driven
  ferroelectricity in van der {Waals} {CuCrP$_2$S$_6$}. Nanoscale.
  2019;\hspace{0pt}11:5163--5170.

\bibitem{zha18}
Zhang~JJ, Lin~L, Zhang~Y, et~al. Type-{II} multiferroic {Hf$_2$VC$_2$F$_2$}
  {MXene} monolayer with high transition temperature. J Am Chem Soc.
  2018;\hspace{0pt}140:9768--9773.

\bibitem{ai19}
Ai~H, Song~X, Qi~S, et~al. Intrinsic multiferroicity in two-dimensional
  {VOCl$_2$} monolayers. Nanoscale. 2019;\hspace{0pt}11:1103--1110.

\bibitem{tan19a}
Tan~H, Li~M, Liu~H, et~al. Two-dimensional ferromagnetic-ferroelectric
  multiferroics in violation of the ${d}^{0}$ rule. Phys Rev B.
  2019;\hspace{0pt}99:195434.

\bibitem{din20}
Ding~N, Chen~J, Dong~S, et~al. Ferroelectricity and ferromagnetism in a
  {$\mathrm{VO}{\mathrm{I}}_{2}$} monolayer: Role of the
  {Dzyaloshinskii-Moriya} interaction. Phys Rev B. 2020;\hspace{0pt}102:165129.

\bibitem{yan17}
Yang~Q, Xiong~W, Zhu~L, et~al. Chemically functionalized phosphorene:
  Two-dimensional multiferroics with vertical polarization and mobile
  magnetism. J Am Chem Soc. 2017;\hspace{0pt}139:11506--11512.

\bibitem{tu19}
Tu~Z, Wu~M. {2D} diluted multiferroic semiconductors upon intercalation. Adv
  Electron Mater. 2019;\hspace{0pt}5:1800960.

\bibitem{zha20}
Zhang~J, Shen~X, Wang~Y, et~al. Design of two-dimensional multiferroics with
  direct polarization-magnetization coupling. Phys Rev Lett.
  2020;\hspace{0pt}125:017601.

\bibitem{duan21}
Duan~X, Huang~J, Xu~B, et~al. A two-dimensional multiferroic metal with
  voltage-tunable magnetization and metallicity. Mater Horiz.
  2021;\hspace{0pt}8:2316--2324.

\bibitem{che14}
Cherifi~RO, Ivanovskaya~V, Phillips~LC, et~al. Electric-field control of
  magnetic order above room temperature. Nat Mater.
  2014;\hspace{0pt}13:345--351.

\bibitem{gon19}
Gong~C, Kim~EM, Wang~Y, et~al. Multiferroicity in atomic van der {Waals}
  heterostructures. Nat Commun. 2019;\hspace{0pt}10:2657.

\bibitem{lu20}
Lu~Y, Fei~R, Lu~X, et~al. Artificial multiferroics and enhanced magnetoelectric
  effect in van der {Waals} heterostructures. ACS Appl Mater Interfaces.
  2020;\hspace{0pt}12:6243--6249.

\bibitem{yan21}
Yang~B, Shao~B, Wang~J, et~al. Realization of semiconducting layered
  multiferroic heterojunctions via asymmetrical magnetoelectric coupling. Phys
  Rev B. 2021;\hspace{0pt}103:L201405.

\bibitem{li21}
Li~P, Zhou~XS, Guo~ZX. Intriguing magnetoelectric effect in two-dimensional
  ferromagnetic/perovskite oxide ferroelectric heterostructure. npj Comput
  Mater. 2022;\hspace{0pt}8:20.

\bibitem{su21}
Su~Y, Li~X, Zhu~M, et~al. Van der {Waals} multiferroic tunnel junctions. Nano
  Lett. 2021;\hspace{0pt}21:175--181.

\bibitem{sha21}
Shang~J, Tang~X, Gu~Y, et~al. Robust magnetoelectric effect in the decorated
  graphene/{In$_2$Se$_3$} heterostructure. ACS Appl Mater Interfaces.
  2021;\hspace{0pt}13:3033--3039.

\bibitem{mat15}
Matsukura~F, Tokura~Y, Ohno~H. Control of magnetism by electric fields. Nat
  Nanotechnol. 2015;\hspace{0pt}10:209--220.

\bibitem{saw10}
Sawicki~M, Chiba~D, Korbecka~A, et~al. Experimental probing of the interplay
  between ferromagnetism and localization in ({Ga},{Mn}){As}. Nat Phys.
  2010;\hspace{0pt}6:22--25.

\bibitem{lee02}
Lee~B, Jungwirth~T, MacDonald~AH. Ferromagnetism in diluted magnetic
  semiconductor heterojunction systems. Semicond Sci Technol.
  2002;\hspace{0pt}17:393--403.

\bibitem{lee02a}
Lee~B, Jungwirth~T, MacDonald~AH. Field-effect magnetization reversal in
  ferromagnetic semiconductor quantum wells. Phys Rev B.
  2002;\hspace{0pt}65:193311.

\bibitem{bou02}
Boukari~H, Kossacki~P, Bertolini~M, et~al. Light and electric field control of
  ferromagnetism in magnetic quantum structures. Phys Rev Lett.
  2002;\hspace{0pt}88:207204.

\bibitem{anh15}
Anh~LD, Hai~PN, Kasahara~Y, et~al. Modulation of ferromagnetism in
  {$(\mathrm{In},\mathrm{Fe})\mathrm{As}$} quantum wells via electrically
  controlled deformation of the electron wave functions. Phys Rev B.
  2015;\hspace{0pt}92:161201.

\bibitem{xin17}
Xing~W, Chen~Y, Odenthal~PM, et~al. Electric field effect in multilayer
  {Cr$_2$Ge$_2$Te$_6$}: a ferromagnetic {2D} material. 2D Mater.
  2017;\hspace{0pt}4:024009.

\bibitem{wan18}
Wang~Z, Zhang~T, Ding~M, et~al. Electric-field control of magnetism in a
  few-layered van der {Waals} ferromagnetic semiconductor. Nat Nanotechnol.
  2018;\hspace{0pt}13:554--559.

\bibitem{sun19}
Sun~YY, Zhu~LQ, Li~Z, et~al. Electric manipulation of magnetism in bilayer van
  der {Waals} magnets. J Phys: Condens Matter. 2019;\hspace{0pt}31:205501.

\bibitem{den18}
Deng~Y, Yu~Y, Song~Y, et~al. Gate-tunable room-temperature ferromagnetism in
  two-dimensional {Fe$_3$GeTe$_2$}. Nature. 2018;\hspace{0pt}563:94--99.

\bibitem{bas83}
Bastard~G, Mendez~EE, Chang~LL, et~al. Variational calculations on a quantum
  well in an electric field. Phys Rev B. 1983;\hspace{0pt}28:3241--3245.

\bibitem{cas07}
Castro~EV, Novoselov~KS, Morozov~SV, et~al. Biased bilayer graphene:
  Semiconductor with a gap tunable by the electric field effect. Phys Rev Lett.
  2007;\hspace{0pt}99:216802.

\bibitem{du21}
Du~L, Hasan~T, Castellanos-Gomez~A, et~al. Engineering symmetry breaking in
  {2D} layered materials. Nat Rev Phys. 2021;\hspace{0pt}3:193--206.

\bibitem{pap99}
Papadakis~SJ, {De Poortere}~EP, Manoharan~HC, et~al. The effect of spin
  splitting on the metallic behavior of a two-dimensional system. Science.
  1999;\hspace{0pt}283:2056--2058.

\bibitem{hab04}
Habib~B, Tutuc~E, Melinte~S, et~al. Negative differential {Rashba} effect in
  two-dimensional hole systems. Appl Phys Lett. 2004;\hspace{0pt}85:3151--3153.

\bibitem{zha09}
Zhang~Y, Tang~TT, Girit~C, et~al. Direct observation of a widely tunable
  bandgap in bilayer graphene. Nature. 2009;\hspace{0pt}459:820--823.

\bibitem{shi15}
Shimazaki~Y, Yamamoto~M, Borzenets~IV, et~al. Generation and detection of pure
  valley current by electrically induced {Berry} curvature in bilayer graphene.
  Nat Phys. 2015;\hspace{0pt}11:1032--1036.

\bibitem{win20}
Winkler~R, Z\"ulicke~U. Collinear orbital antiferromagnetic order and
  magnetoelectricity in quasi-two-dimensional itinerant-electron paramagnets,
  ferromagnets, and antiferromagnets. Phys Rev Research.
  2020;\hspace{0pt}2:043060.

\bibitem{gor94}
Gorbatsevich~AA, Kapaev~VV, Kopaev~YV. Magnetoelectric phenomena in
  nanoelectronics. Ferroelectrics. 1994;\hspace{0pt}161:303--310.

\bibitem{zha19}
Zhang~D, Shi~M, Zhu~T, et~al. Topological axion states in the magnetic
  insulator {${\mathrm{MnBi}}_{2}{\mathrm{Te}}_{4}$} with the quantized
  magnetoelectric effect. Phys Rev Lett. 2019;\hspace{0pt}122:206401.

\bibitem{otr19}
Otrokov~MM, Rusinov~IP, Blanco-Rey~M, et~al. Unique thickness-dependent
  properties of the van der {Waals} interlayer antiferromagnet
  {${\mathrm{MnBi}}_{2}{\mathrm{Te}}_{4}$} films. Phys Rev Lett.
  2019;\hspace{0pt}122:107202.

\bibitem{li19}
Li~J, Li~Y, Du~S, et~al. Intrinsic magnetic topological insulators in van der
  {Waals} layered {MnBi$_2$Te$_4$}-family materials. Sci Adv.
  2019;\hspace{0pt}5:eaaw5685.

\bibitem{liu20}
Liu~C, Wang~Y, Li~H, et~al. Robust axion insulator and {Chern} insulator phases
  in a two-dimensional antiferromagnetic topological insulator. Nat Mater.
  2020;\hspace{0pt}19:522--527.

\bibitem{gao21}
Gao~A, Liu~YF, Hu~C, et~al. Layer {Hall} effect in a {2D} topological axion
  antiferromagnet. Nature. 2021;\hspace{0pt}595:521--525.

\bibitem{zhu21}
Zhu~T, Wang~H, Zhang~H, et~al. Tunable dynamical magnetoelectric effect in
  antiferromagnetic topological insulator {MnBi$_2$Te$_4$} films. npj Comput
  Mater. 2021;\hspace{0pt}7:121.

\bibitem{zue14}
Z\"ulicke~U, Winkler~R. Magnetoelectric effect in bilayer graphene controlled
  by valley-isospin density. Phys Rev B. 2014;\hspace{0pt}90:125412.

\bibitem{kam19}
Kammermeier~M, Wenk~P, Z\"ulicke~U. In-plane magnetoelectric response in
  bilayer graphene. Phys Rev B. 2019;\hspace{0pt}100:075421.

\bibitem{gon13}
Gong~Z, Liu~GB, Yu~H, et~al. Magnetoelectric effects and valley-controlled spin
  quantum gates in transition metal dichalcogenide bilayers. Nat Commun.
  2013;\hspace{0pt}4:2053.

\bibitem{sch16}
Schaibley~JR, Yu~H, Clark~G, et~al. Valleytronics in {2D} materials. Nat Rev
  Mater. 2016;\hspace{0pt}1:16055.

\bibitem{som00}
Sometani~T. Image method for a dielectric plate and a point charge. Eur J Phys.
  2000;\hspace{0pt}21:549--554.

\bibitem{kum89}
Kumagai~M, Takagahara~T. Excitonic and nonlinear-optical properties of
  dielectric quantum-well structures. Phys Rev B.
  1989;\hspace{0pt}40:12359--12381.

\bibitem{dir31}
Dirac~PAM. Quantised singularities in the electromagnetic field. Proc R Soc
  Lond A. 1931;\hspace{0pt}133:60--72.

\bibitem{sch69}
Schwinger~J. A magnetic model of matter. Science.
  1969;\hspace{0pt}165:757--761.

\bibitem{wit79}
Witten~E. Dyons of charge $e\theta$/$2\pi$. Phys Lett B.
  1979;\hspace{0pt}86:283--287.

\bibitem{pla21}
Planelles~J. Axion electrodynamics in topological insulators for beginners;
  2021. \urlprefix\url{https://arxiv.org/abs/2111.07290}.

\bibitem{som49}
Sommerfeld~A. Partial differential equations in physics. (Lectures on
  Theoretical Physics; Vol.~VI). New York (NY): Academic Press; 1949.

\bibitem{che99}
Chew~WC. Waves and fields in inhomogenous media. Piscataway (NJ): Wiley-IEEE
  Press; 1999. IEEE Press Series on Electromagnetic Wave Theory.

\bibitem{cai13}
Cai~W. Computational methods for electromagnetic phenomena: Electrostatics in
  solvation, scattering, and electron transport. Cambridge: Cambridge
  University Press; 2013.

\bibitem{fon63}
Foner~S. High-field antiferromagnetic resonance in {Cr$_2$O$_3$}. Phys Rev.
  1963;\hspace{0pt}130:183--197.

\bibitem{lal67}
Lal~HB, Srivastava~R, Srivastava~KG. Magnetoelectric effect in {Cr$_2$O$_3$}
  single crystal as studied by dielectric-constant method. Phys Rev.
  1967;\hspace{0pt}154:505--507.

\bibitem{wie94}
Wiegelmann~H, Jansen~AGM, Wyder~P, et~al. Magnetoelectric effect of
  {Cr$_2$O$_3$} in strong static magnetic fields. Ferroelectrics.
  1994;\hspace{0pt}162:141--146.

\bibitem{asc74}
Ascher~E. Kineto-electric and kinetomagnetic effects in crystals. Int J
  Magnetism. 1974;\hspace{0pt}5:287--295.

\bibitem{bro68a}
Brown~WF, Hornreich~RM, Shtrikman~S. Upper bound on the magnetoelectric
  susceptibility. Phys Rev. 1968;\hspace{0pt}168:574--577.

\end{thebibliography}

%Note that if the \verb"endfloat" package is used on a document containing any appendices, the \verb"\processdelayedfloats" command must be included immediately before the \verb"\appendix" command in order to ensure that the floats belonging to the main body of the text are numbered as such.

%\processdelayedfloats %%% See above for an explanation of why this command might be needed here.

\appendix

\section{Conventions for magnetoelectric constitutive relations}
\label{app:ConstRel}

Depending on circumstances, different forms of constitutive
relations for the responses to electromagnetic fields in a
magnetoelectric medium are preferred~\cite{riv09}. If the
electric field $\vek{\Ec}$ and the magnetic field $\vek{\Bc}$
are adopted as controllable quantities, the form
\begin{subequations}\label{eqs:EBconst}
\begin{eqnarray}
\vek{\Pc} &=& \underline{\chi}_\Ec\, \vek{\Ec} +
\underline{\alpha}\, \vek{\Bc}\,\, , \\ \label{eq:MfromE}
\vek{\Mc} &=& \underline{\chi}_\Bc\, \vek{\Bc} +
\underline{\alpha}^\mathrm{T} \, \vek{\Ec}
\end{eqnarray}
\end{subequations}
is most suitable. Here $\vek{\Pc}\equiv -\partial F/\partial
\vek{\Ec}$ and $\vek{\Mc}\equiv -\partial F/\partial\vek{\Bc}$
are the electric polarization and the magnetization,
respectively, given in terms of derivatives of the
magnetoelectric material's free energy per unit volume $F$. See
Eq.~(\ref{eq:freeEB}). Alternatively, in situations when
$\vek{\Hc}\equiv \mu_0^{-1}\, \vek{\Bc} - \vek{\Mc}$ is fixed
instead of $\vek{\Bc}$, it is more practical to use the
relations
\begin{subequations}\label{eqs:EHconst}
\begin{eqnarray}
\vek{\Pc} &=& \underline{\chi}^\prime_\Ec\, \vek{\Ec} +
\underline{\alpha}^\prime\, \vek{\Hc} \,\, , \\
\vek{\Mc} &=& \underline{\chi}_\Hc\, \vek{\Hc} + \mu_0^{-1}\,
\underline{\alpha}^{\prime\,\mathrm{T}}\, \vek{\Ec} \,\, .
\end{eqnarray}
\end{subequations}
The derived fields $\vek{\Dc}\equiv \epsilon_0\, \vek{\Ec} +
\vek{\Pc}$ and $\vek{\Hc}$ can then be expressed in terms of the
fundamental fields $\vek{\Ec}$ and $\vek{\Bc}$ equally well
using either one of the two related definitions
$\underline{\alpha}$ and $\underline{\alpha}^\prime$ of the
magnetoelectric tensor;
\begin{subequations}
\begin{eqnarray}
\vek{\Dc} &=& \underline{\epsilon}\, \vek{\Ec} +
\underline{\alpha}\, \vek{\Bc} \,\,\, \equiv\,\,\,
\underline{\epsilon} \, \vek{\Ec} + \underline{\alpha}^\prime\,
\underline{\mu}^{-1}\, \vek{\Bc} \,\, , \\ \vek{\Hc} &=&
\underline{\mu}^{-1}\, \vek{\Bc} - \underline{\alpha}^\mathrm{T}
\, \vek{\Ec}\,\,\, \equiv\,\,\, \underline{\mu}^{-1} \big(
\vek{\Bc} - \underline{\alpha}^{\prime\,\mathrm{T}}\,\vek{\Ec}
\big) \,\, ,
\end{eqnarray}
\end{subequations}
with $\underline{\epsilon} = \epsilon_0\, \mathbb{1} +
\underline{\chi}_\Ec$ and $\underline{\mu} = \mu_0\, \big(
\mathbb{1} + \underline{\chi}_\Hc \big) \equiv \big( \mu_0^{-1}
\, \mathbb{1} - \underline{\chi}_\Bc\big)^{-1}$. Similarly, the
constitutive relations that give $\vek{\Dc}$ and $\vek{\Bc}$ in
terms of $\vek{\Ec}$ and $\vek{\Hc}$ read (with
$\underline{\epsilon}^\prime = \epsilon_0\, \mathbb{1} +
\underline{\chi}^\prime_\Ec$)
\begin{subequations}
\begin{eqnarray}
\vek{\Dc} &=& \underline{\epsilon}^\prime\, \vek{\Ec} +
\underline{\alpha}^\prime\, \vek{\Hc} \,\,\, \equiv\,\,\,
\underline{\epsilon}^\prime \, \vek{\Ec} + \underline{\alpha}\,
\underline{\mu}\, \vek{\Hc} \,\, , \\
\vek{\Bc} &=& \underline{\mu}\, \vek{\Hc} +
\underline{\alpha}^{\prime\,\mathrm{T}}\, \vek{\Ec}\,\,\,
\equiv\,\,\, \underline{\mu}\, \big( \vek{\Hc} +
\underline{\alpha}^\mathrm{T}\,\vek{\Ec} \big) \,\, .
\end{eqnarray}
\end{subequations}

The susceptibilities $\underline{\chi}_\Ec$,
$\underline{\chi}_\Bc$, $\underline{\chi}_\Hc$, and by extension
$\underline{\epsilon}$ and $\underline{\mu}$, are symmetric
tensors~\cite{nye57}, but $\underline{\alpha}$ and
$\underline{\alpha}^\prime$ are not necessarily
symmetric~\cite{ode70,lan84}. The necessity to distinguish
$\underline{\chi}^\prime_\Ec$ from $\underline{\chi}_\Ec$ arises
in a magnetoelectric medium~\cite{asc74} because the condition
$\vek{\Hc}=\vek{0}$ at finite $\vek{\Ec}$ requires finite
$\vek{\Bc}\equiv\underline{\alpha}^{\prime\,\mathrm{T}}\,
\vek{\Ec}$, and the latter makes a contribution to $\vek{\Pc}$.
More specifically, $\underline{\chi}^\prime_\Ec =
\underline{\chi}_\Ec + \underline{\alpha}\, \underline{\mu}\,
\underline{\alpha}^\mathrm{T} \equiv \underline{\chi}_\Ec +
\underline{\alpha}^\prime\, \underline{\mu}^{-1}\,
\underline{\alpha}^{\prime\,\mathrm{T}}$. Hence, in principle,
the meaning of the dielectric tensor in a magnetoelectric
medium needs to be carefully defined based on the physical
situation and/or details of the measurement protocol. In
contrast, there is no ambiguity associated with
$\underline{\mu}$ or the magnetoelectric tensor
$\underline{\alpha}\equiv \underline{\alpha}^\prime\,
\underline{\mu}^{-1}$. Practically, the fundamental
limit~\cite{bro68a} $\big|(\underline{\alpha})_{ij}\big| \le
\sqrt{\big(\underline{\chi}_\Ec\big)_{ii}\, \big(
\underline{\chi}_\Bc \big)_{jj}}$ on the magnitude of
magnetoelectric-tensor elements also constrains the difference
between $\underline{\chi}^\prime_\Ec$ and
$\underline{\chi}_\Ec$. For example, in a uniaxial
magnetoelectric medium where $(\underline{\mu})_{ij}=\mu_{ii}\,
\delta_{ij}$, $(\underline{\alpha})_{ij} = \alpha_{ii}\,
\delta_{ij}$ etc., the inequality
\begin{equation}
\frac{\big| \big( \underline{\chi}^\prime_\Ec -
\underline{\chi}_\Ec \big)_{ii} \big|}{\big(
\underline{\chi}_\Ec \big)_{ii}} \le
\frac{(\underline{\mu})_{ii}}{\mu_0} - 1
\end{equation}
holds whose r.h.s.\ is typically small. Moreover, the
magnitudes of magnetoelectric-tensor elements in real materials
are generally well below their maximum value mandated by
thermodynamic stability~\cite{bro68a}.

\section{Formalism for determining image dyons}

\subsection{Case of a semi-infinite magnetoelectric medium}
\label{app:semiinf}

\subsubsection{Electric point charge as the source}

To discuss the case where an electric charge is outside a
semi-infinite magnetoelectric medium, we assume the region $z>0$
($z<0$) to be vacuum (a magnetoelectric) as illustrated in
Fig.~\ref{fig:qSemi}(a). We introduce specific Ans{\"a}tze
for the scalar potentials;
\begin{subequations}\label{eq:AnsatzOutSemi}
\begin{eqnarray}
V(R, \theta, z>0) &=& \frac{1}{4\pi\epsilon_0}\left[
\frac{q_0}{\sqrt{(z-z_0)^2+R^2}} + \frac{q^{-}_1}{\sqrt{(z+
z_0)^2+R^2}}\right] \quad , \\
V(R, \theta, z<0) &=& \frac{1}{4\pi\epsilon_0}\left[
\frac{q_0+q^{+}_1}{\sqrt{(z-z_0)^2+R^2}} \right] \quad , \\
U(R, \theta, z>0) &=& \frac{\mu_0}{4\pi}\left[
\frac{g^{-}_1}{\sqrt{(z+z_0)^2+R^2}} \right] \quad , \\
U(R, \theta, z<0) &=& \frac{\mu_0}{4\pi}\left[
\frac{g^{+}_1}{\sqrt{(z-z_0)^2+r^2}} \right] \quad .
\end{eqnarray}
\end{subequations}
Applying boundary conditions as specified by Eq.~\eqref{eq:BC},
we find the image charges and monopoles as
\begin{subequations}\label{eq:ChargeOutSemi}
\begin{eqnarray}
q^{+}_1 = q^{-}_1 &=& -q_0\, \frac{(\epsilon - \epsilon_0)
(1/\mu + 1/\mu_0) + \alpha^2}{(\epsilon + \epsilon_0) (1/\mu +
1/\mu_0) + \alpha^2} \quad , \\
g^{+}_1 = -g^{-}_1 &=& \frac{q_0}{\mu_0}\, \frac{2\alpha}{(
\epsilon + \epsilon_0)(1/\mu + 1/\mu_0) + \alpha^2} \quad .
\end{eqnarray}
\end{subequations}

Similarly, when the source charge is inside the magnetoelectric
medium and located at $(0,0,-z_0)$, where $z_0>0$ [see
Fig.~\ref{fig:qSemi}(b)], we use the Ans{\"a}tze
\begin{subequations}
\begin{eqnarray}
V(R, \theta, z>0) &=& \frac{1}{4\pi\epsilon} \left[ \frac{q_0 +
q^{-}_1}{\sqrt{(z + z_0)^2 + R^2}} \right] \quad , \\
V(R, \theta, z<0) &=& \frac{1}{4\pi\epsilon} \left[
\frac{q_0}{\sqrt{(z + z_0)^2 + R^2}} + \frac{q^{+}_1}{\sqrt{(z -
z_0)^2 + R^2}} \right] \quad , \\
U(R, \theta, z>0) &=& \frac{\mu}{4\pi} \left[
\frac{g^{-}_1}{\sqrt{(z + z_0)^2 + R^2}} \right] \quad , \\
U(R, \theta, z<0) &=& \frac{\mu}{4\pi} \left[
\frac{g^{+}_1}{\sqrt{(z - z_0)^2 + R^2}} \right] \quad .
\end{eqnarray}
\end{subequations}
The image charges satisfying the boundary conditions are found
as
\begin{subequations}\label{eq:ChargeInSemiME}
\begin{eqnarray}
q^{+}_1 = q^{-}_1 &=& -q_0\, \frac{(-\epsilon +\epsilon_0)
(1/\mu + 1/\mu_0) + \alpha^2}{(\epsilon +\epsilon_0) (1/\mu +
1/\mu_0)+\alpha^2} \quad , \\
g^{+}_1=-g^{-}_1 &=& \frac{q_0}{\mu_0}\, \frac{2\alpha}{(
\epsilon + \epsilon_0) (1/\mu + 1/\mu_0) + \alpha^2} \quad .
\end{eqnarray}
\end{subequations}
Note that the image charges from Eq.~\eqref{eq:ChargeInSemiME}
are the same as the ones in Eg.~\eqref{eq:ChargeOutSemi} if we
simultaneously swap $\epsilon$ with $\epsilon_0$ and $\mu$ with
$\mu_0$. Also, in the limit $\alpha = 0$, we recover the known
results for the situation where an electric point charge is
near/inside a semi-infinite dielectric medium.

\subsubsection{Dyon as the source}
\label{app:dyonsource}

To assist the calculation for the cases where the
magnetoelectric medium is a slab with finite width, it is useful
to obtain the results when the source is a dyon, i.e., a point
particle located at $\rr_\mathrm{d}=(0,0, \pm z_{\rm d})$ having
both magnetic and electric charges $g_{\rm d}$ and $q_{\rm d}$,
respectively. Considering this is needed as part of our
iterative process for satisfying boundary conditions at
interfaces with magnetoelectrics. 

If we started with the upper boundary, the dyon is always on the
side of the magnetoelectric medium. Assuming that the region
$z<0$ is where the dyon is located, the Ans{\"a}tze are
\begin{subequations}
\begin{eqnarray}
V(R, \theta, z>0) &=& \frac{1}{4\pi\epsilon'} \left[
\frac{q_\mathrm{d} + q^{-}_1}{\sqrt{(z + z_\mathrm{d})^2 + R^2}}
\right] \quad , \\
V(R, \theta, z<0) &=& \frac{1}{4\pi\epsilon'}\left[
\frac{q_\mathrm{d}}{\sqrt{(z + z_\mathrm{d})^2 + R^2}} +
\frac{q^{+}_1}{\sqrt{(z - z_\mathrm{d})^2 + R^2}} \right]
\quad , \\
U(R, \theta, z>0) &=& \frac{\mu'}{4\pi}\left[ \frac{g_\mathrm{d}
+ g^{-}_1}{\sqrt{(z + z_\mathrm{d})^2 + R^2}} \right] \quad , \\
U(R, \theta, z<0) &=& \frac{\mu'}{4\pi}\left[
\frac{g_\mathrm{d}}{\sqrt{(z + z_\mathrm{d})^2 + R^2}} +
\frac{g^{+}_1}{\sqrt{(z - z_\mathrm{d})^2 + R^2}} \right]\quad .
\end{eqnarray}
\end{subequations}
Depending on whether the original electric source charge
$q_0$ is outside or inside the magnetoelectric slab, we choose
$\epsilon_0$ ($\mu_0$) or $\epsilon$ ($\mu$) for $\epsilon'$
($\mu'$), respectively, to keep with our overall convention. 

The resulting image charges for the case when the original
source charge $q_0$ is outside the magnetoelectric slab are
\begin{subequations}\label{eq:ChargeOutGeneral}
\begin{eqnarray}
q^{+}_1 = q^{-}_1 &=& \frac{-q_\mathrm{d} [( -\epsilon +
\epsilon_0) (1/\mu + 1/\mu_0) + \alpha^2] + g_\mathrm{d} (2
\alpha \epsilon_0\mu_0/\mu)}{(\epsilon + \epsilon_0) (1/\mu +
1/\mu_0)+\alpha^2} \quad , \\
g^{+}_1 = -g^{-}_1 &=& \frac{q_\mathrm{d} (2 \alpha \epsilon /
\mu_0 \epsilon_0) + g_\mathrm{d} [( \epsilon +\epsilon_0) (1/\mu
- 1/\mu_0) + \alpha^2]}{[(\epsilon + \epsilon_0) (1/\mu +
1/\mu_0) +\alpha^2]} \quad .
\end{eqnarray}
\end{subequations}
Alternatively, when the original source charge $q_0$ is inside
the magnetoelectric slab, the image charges are
\begin{subequations}\label{eq:ChargeInGeneral}
\begin{eqnarray}
q^{+}_1 = q^{-}_1 &=& \frac{   -q_\mathrm{d} [( -\epsilon +
\epsilon_0) (1/\mu + 1/\mu_0) + \alpha^2] + g_\mathrm{d} (2
\alpha \epsilon)}{(\epsilon + \epsilon_0) (1/\mu + 1/\mu_0) +
\alpha^2} \quad , \\
g^{+}_1 = -g^{-}_1 &=& \frac{q_\mathrm{d}(2\alpha/\mu) +
g_\mathrm{d} [( \epsilon + \epsilon_0) (1/\mu - 1/\mu_0) +
\alpha^2]}{[(\epsilon + \epsilon_0)(1/\mu + 1/\mu_0) +
\alpha^2]} \quad .
\end{eqnarray}
\end{subequations}

\subsection{Case of a finite-width magnetoelectric slab}
\label{app:finite}

\subsubsection{Electric point charge located outside the slab}

When $z_0>0$, the Ans{\"a}tze are
\begin{subequations}\label{eq:ansatzQaboveSlab}
\begin{eqnarray}
V(R, \theta, z>0) &=& \frac{1}{2\pi^2 \epsilon_0}
\int^{\infty}_0 \int^{\infty}_0 d\gamma\, d\eta\,\,
\frac{\cos{(\gamma R)}}{u} \left( q_0\, \mathrm{e}^{-u|z-z_0|}
+ A\, \mathrm{e}^{-u z} \right) \,\, , \nonumber \\ \\
V(R, \theta, 0>z>-w) &=& \frac{1}{2\pi^2\epsilon_0}
\int^{\infty}_0 \int^{\infty}_0 d\gamma\, d\eta\,\,
\frac{\cos{(\gamma R)}}{u} \left( B\, \mathrm{e}^{u z} + C\,
\mathrm{e}^{-u z}\right) \quad , \\
V(R, \theta, z<-w) &=& \frac{1}{2\pi^2\epsilon_0}
\int^{\infty}_0 \int^{\infty}_0 d\gamma\, d\eta\,\,
\frac{\cos{(\gamma R)}}{u}\,\, D\, \mathrm{e}^{uz} \quad , \\
U(R, \theta, z>0) &=& \frac{\mu_0}{2\pi^2}\int^{\infty}_0
\int^{\infty}_0 d\gamma\, d\eta\,\, \frac{\cos{(\gamma R)}}{u}
\,\, E\, \mathrm{e}^{-uz} \quad , \\
U(R, \theta, 0>z>-w) &=& \frac{\mu_0}{2\pi^2} \int^{\infty}_0
\int^{\infty}_0 d\gamma\, d\eta\,\, \frac{\cos{(\gamma R)}}{u}
\left( F\, \mathrm{e}^{uz} + G\, \mathrm{e}^{-uz} \right)
\quad , \\
U(R, \theta, z<-w) &=& \frac{\mu_0}{2\pi^2} \int^{\infty}_0
\int^{\infty}_0 d\gamma\, d\eta\,\, \frac{\cos{(\gamma R)}}{u}
\,\, H\, \mathrm{e}^{uz} \quad .
\end{eqnarray}
\end{subequations}  
Here we absorbed the source-charge terms into $B$, $C$ and $D$
in the Ans{\"a}tze for $V(R, \theta, 0>z>-w)$ and $V(R, \theta,
z<-w)$. Below we list the solutions when $\epsilon_0=\epsilon$
and $\mu_0=\mu$ as an example, i.e., when the only difference
between the regions inside and outside the slab is the finite
magnetoelectricity coupling inside.
\begin{equation}
\begin{split}
A&=q_0\, \frac{-{\rm e}^{-u z_0}\sigma +{\rm
e}^{-u(z_0+2w)}\sigma}{1-{\rm e}^{-2uw}\sigma},\\
B&=q_0\, \frac{4\epsilon_0}{4\epsilon_0+\alpha^2 \mu_0}
\frac{{\rm e}^{-uz_0}} {1-{\rm e}^{-2uw}\sigma}, \qquad
C=0,\\
D&=B,\\
E&=q_0\, \frac{2\alpha}{4\epsilon_0+\alpha^2 \mu_0}
\frac{-{\rm e}^{-u z_0} +{\rm e}^{-u(z_0+2w)}}{1-{\rm
e}^{-2uw}\sigma},\\
F&=q_0\, \frac{2\alpha}{4\epsilon_0+\alpha^2 \mu_0}
\frac{{\rm e}^{-uz_0}}{1-{\rm e}^{-2uw}\sigma},\qquad
G=q_0\, \frac{2\alpha}{4\epsilon_0+\alpha^2 \mu_0}
\frac{{\rm e}^{-u(z_0+2w)}}{1-{\rm e}^{-2uw}\sigma},\\
H&=0,
\end{split}
\end{equation}
with $\sigma=\frac{\alpha^2 \mu_0}{4\epsilon_0+\alpha^2 \mu_0}$.
We can interpret the above results in terms of image charges
shown in Table~\ref{tab:SlabQabove} by applying the following
relation:
\begin{equation}
\frac{1}{1-{\rm e}^{-2uw}\sigma}=\sum^{\infty}_{k=1}{\rm
e}^{-2kuw}\sigma^{k-1}.
\end{equation}

From expressions given in Table~\ref{tab:SlabQabove}, we can see
that, when $\epsilon_0=\epsilon$ and $\mu_0=\mu$,  $V(R, \theta,
0>z>-w)$ and $V(R, \theta, z<-w)$ share the same expression, and
$U(R, \theta, z<-w)=0$. Thus, the electric field is continuous
and $\vek{\Bc}=0$ in the region where $z<0$, as shown in
Fig.~\ref{fig:QoutSlab}.
 
\begin{table}
    \tbl{The potentials in terms of image charges when the point charge is above the magnetoelectric slab. Note that  $d^{\rm{L-}}_{1}$ does not exist by construction.}
    {\begin{tabular}{|c|c|c|}
    \hline
    &Potential& \makecell{Image charge\\ when $\epsilon_0=\epsilon$ and $\mu_0=\mu$}
    \\\hline
    $V(R, \theta, z>0)$
    &$\displaystyle
    \frac{1}{4\pi\epsilon_0}
    \left[
    \frac{q_0}{|\rr-\rr_{0}|}+\sum_{m={\rm U,L}}\sum^{\infty}_{k=1}\frac{q^{m-}_{k}}{|\rr-\rr^{-}_{k}|}
    \right]$
    & \multirow{3}{*}[-0.8em]{\makecell{
    $q^{\rm{L\pm}}_{k}=0$\\
    \\
    $q^{\rm{U\pm}}_{k}=
    \begin{cases}
    -q_0\sigma&k=1\\
    q_0(1-\sigma)\sigma^{(k-1)}&k\geq2    
    \end{cases}$
    }}
    \\\cline{1-2}
    $V(R, \theta, 0>z>-w)$
    &$\displaystyle
    \frac{1}{4\pi\epsilon_0}
    \left[
    \frac{q_0}{|\rr-\rr_{0}|}+\sum^{\infty}_{k=1}\left(
    \frac{q^{\rm{U}+}_{k}}{|\rr-\rr^{+}_{k}|}+\frac{q^{\rm{L-}}_{k}}{|\rr-\rr^{\rm{-}}_{k}|}
    \right)
    \right]$
    &    
    \\\cline{1-2}
    $V(R, \theta, z<-w)$
    &$\displaystyle
    \frac{1}{4\pi\epsilon_0}
    \left[
    \frac{q_0}{|\rr-\rr_{0}|}
    +\sum_{m={\rm U,L}}\sum^{\infty}_{k=1}\frac{q^{m+}_{k}}{|\rr-\rr^{+}_{k}|}
    \right]$
    &
    \\\hline
    $U(R, \theta, z>0)$
    &$\displaystyle\frac{\mu_0}{4\pi}
    \sum_{m={\rm U,L}}\sum^{\infty}_{k=1}\frac{g^{m-}_{k}}{|\rr-\rr^{-}_{k}|}$
    &\multirow{3}{*}[-1.5em]{
    $\begin{aligned}
    &g^{\rm{U+}}_{k}=-g^{\rm{U-}}_{k}\\
    =&-g^{\rm{L+}}_{k}=g^{\rm{L-}}_{k+1}\\
    =&q_0\frac{2\alpha}{4\epsilon_0+\alpha^2 \mu_0}\sigma^{k-1}
    \end{aligned}$
    }
    \\\cline{1-2}
    $U(R, \theta, 0>z>-w)$
    &$\displaystyle \frac{\mu_0}{4\pi}
    \sum^{\infty}_{k=1}\left(
    \frac{g^{\rm{U+}}_{k}}{|\rr-\rr^{\rm{+}}_{k}|}+\frac{g^{\rm{L-}}_{k}}{|\rr-\rr^{\rm{-}}_{k}|}
    \right)$
    &
    \\\cline{1-2}
    $U(R, \theta, z<-w)$
    &$\displaystyle \frac{\mu_0}{4\pi}
    \sum_{m={\rm U,L}}\sum^{\infty}_{k=1}\frac{g^{m+}_{k}}{|\rr-\rr^{+}_{k}|}$
    &\\\hline
    \end{tabular}}
    \label{tab:SlabQabove}
\end{table}

\begin{table}
\tbl{The potentials in terms of image charges when the point
charge is in the magnetoelectric slab. Note that $d^{m-}_{-1}$,
$d^{\rm{U+}}_{-1}$ and $d^{\rm{L-}}_{1}$ do not exist by
construction.}
{\begin{tabular}{|c|c|c|}
\hline
&Potential& Image charge when $\epsilon_0=\epsilon$ and $\mu_0=\mu$\\\hline
$V(R, \theta, z>0)$
        &$\displaystyle
        \frac{1}{4\pi\epsilon}
        \left[
        \frac{q_0}{|\rr-\rr_{0}|}+\sum_{m={\rm U,L}}\sum^{\infty}_{k=1}\frac{q^{m-}_{\pm k}}{|\rr-\rr^{-}_{\pm k}|}
        \right]$
        &\multirow{6}{*}{
        $\begin{aligned}
        &q^{\rm{U+}}_{k}=-q^{\rm{U+}}_{-(k+1)}\\
        =&q^{\rm{L+}}_{-k}=-q^{\rm{L+}}_{k}=-q_0\sigma^{k};\\
        &q^{\rm{L-}}_{-(k+1)}=-q^{\rm{L-}}_{k+1}\\
        =&q^{\rm{U-}}_{k}=-q^{\rm{U-}}_{-(k+1)}=-q_0\sigma^{k-1}\\
        \\
        \\
        &g^{\rm{U+}}_{-k}=g^{\rm{L+}}_{k}\\
        =&g^{\rm{L-}}_{k}=g^{\rm{U-}}_{-k}=0;\\
        &g^{\rm{U+}}_{k}=-g^{\rm{L+}}_{-k}\\
        =&g^{\rm{L}-}_{-(k+1)}=-g^{\rm{U-}}_{k}=q_0\frac{2\alpha}{4\epsilon_0+\alpha^2 \mu_0}\sigma^{k-1}
        \end{aligned}$
        }
        \\\cline{1-2}
        $V(R, \theta, 0>z>-w)$
        &$\displaystyle
        \frac{1}{4\pi\epsilon}
        \left[
        \frac{q_0}{|\rr-\rr_{0}|}+\sum^{\infty}_{k=1}\left(
        \frac{q^{\rm{U+}}_{\pm k}}{|\rr-\rr^{\rm{+}}_{\pm k}|}+\frac{q^{\rm{L-}}_{\pm k}}{|\rr-\rr^{\rm{-}}_{\pm k}|}
        \right)
        \right]$
        &
        \\\cline{1-2}
        $V(R, \theta, z<-w)$
        &$\displaystyle
        \frac{1}{4\pi\epsilon}
        \left[
        \frac{q_0}{|\rr-\rr_{0}|}
        +\sum_{m={\rm U,L}}\sum^{\infty}_{k=1}\frac{q^{m+}_{\pm k}}{|\rr-\rr^{+}_{\pm k}|}
        \right]$
        &
        \\\hline
        $U(R, \theta, z>0)$&$\displaystyle \frac{\mu}{4\pi}
        \sum_{m={\rm U,L}}\sum^{\infty}_{k=1}\frac{g^{m-}_{\pm k}}{|\rr-\rr^{-}_{\pm k}|}$
        &
        \\ \cline{1-2}
        $U(R, \theta, 0>z>-w)$&$\displaystyle \frac{\mu}{4\pi}
        \sum^{\infty}_{k=1}\left(
        \frac{g^{\rm{U+}}_{\pm k}}{|\rr-\rr^{\rm{+}}_{\pm k}|}+\frac{g^{\rm{L-}}_{\pm k}}{|\rr-\rr^{\rm{-}}_{\pm k}|}
        \right)$
        &
        \\\cline{1-2}
        $U(R, \theta, z<-w)$&$\displaystyle \frac{\mu}{4\pi}
        \sum_{m={\rm U,L}}\sum^{\infty}_{k=1}\frac{g^{m+}_{\pm k}}{|\rr-\rr^{+}_{\pm k}|}$
        &
        \\ \hline
        \end{tabular}}
        \label{tab:SlabQin}
    \end{table}

\subsubsection{Electric point harge located inside the slab}

When $z_0>0$ and $\rr_0=(0,0,-z_0)$, the Ans{\"a}tze are
\begin{subequations}
\begin{eqnarray}
V(R, \theta, z>0) &=& \frac{1}{2\pi^2\epsilon} \int^{\infty}_0
\int^{\infty}_0 d\gamma\, d\eta\,\, \frac{\cos{(\gamma R)}}{u}
\,\, A\, \mathrm{e}^{-uz} \quad , \\
V(R, \theta, 0>z>-w) &=& \frac{1}{2\pi^2\epsilon}
\int^{\infty}_0 \int^{\infty}_0 d\gamma\, d\eta\,\,
\frac{\cos{(\gamma R)}}{u} \left( q_0\, \mathrm{e}^{-u|z+z_0|}
+ B\, \mathrm{e}^{uz} + C\, \mathrm{e}^{-uz} \right) \,\, ,
\nonumber \\ \\
V(R, \theta, z<-w) &=& \frac{1}{2\pi^2\epsilon} \int^{\infty}_0
\int^{\infty}_0 d\gamma\, d\eta\,\, \frac{\cos{(\gamma R)}}{u}
\,\, D\, \mathrm{e}^{uz} \quad , \\
U(R, \theta, z>0) &=& \frac{\mu}{2\pi^2} \int^{\infty}_0
\int^{\infty}_0 d\gamma\, d\eta\,\, \frac{\cos{(\gamma R)}}{u}
\,\, E\, \mathrm{e}^{-uz} \quad , \\
U(R, \theta, 0>z>-w) &=& \frac{\mu}{2\pi^2} \int^{\infty}_0
\int^{\infty}_0 d\gamma\, d\eta\,\, \frac{\cos{(\gamma R)}}{u}
\left( F\, \mathrm{e}^{uz} + G\, \mathrm{e}^{-uz} \right)
\quad , \\
U(R, \theta, z<-w) &=& \frac{\mu}{2\pi^2} \int^{\infty}_0
\int^{\infty}_0 d\gamma\, d\eta\,\, \frac{\cos{(\gamma R)}}{u}
\,\, H\, \mathrm{e}^{uz} \quad .
\end{eqnarray}
\end{subequations}
Here we absorbed the source-charge terms into $A$ and $D$ in the
Ans{\"a}tze for $V(R, \theta, z>0)$ and $V(R, \theta, z<-w)$.
Again, we list the solutions when $\epsilon_0=\epsilon$ and
$\mu_0=\mu$ as an example:
\begin{equation}
\begin{split}
A&=q_0\, \frac{4\epsilon_0}{4\epsilon_0+\alpha^2 \mu_0}
\frac{{\rm e}^{-uz_0}}{1-{\rm e}^{-2uw}\sigma},\\
B&=q_0\, \frac{-{\rm e}^{-u z_0}\sigma +{\rm
e}^{u(z_0-2w)}\sigma}{1-{\rm e}^{-2uw}\sigma},\qquad
C=q_0\, \frac{{\rm e}^{-u (z_0+2w)}\sigma -{\rm
e}^{-u(-z_0+2w)}\sigma}{1-{\rm e}^{-2uw}\sigma},\\
D&=q_0\, \frac{4\epsilon_0}{4\epsilon_0+\alpha^2 \mu_0}
\frac{{\rm e}^{uz_0}}{1-{\rm e}^{-2uw}\sigma},\\
E&=q_0\, \frac{2\alpha}{4\epsilon_0+\alpha^2 \mu_0}
\frac{-{\rm e}^{-u z_0}+{\rm e}^{-u(-z_0+2w)}}{1-{\rm
e}^{-2uw}\sigma},\\
F&=q_0\, \frac{2\alpha}{4\epsilon_0+\alpha^2 \mu_0}
\frac{{\rm e}^{-uz_0}}{1-{\rm e}^{-2uw}\sigma},\qquad
G=q_0\, \frac{2\alpha}{4\epsilon_0+\alpha^2 \mu_0}
\frac{{\rm e}^{-u(-z_0+2w)}}{1-{\rm e}^{-2uw}\sigma},\\
H&=q_0\, \frac{2\alpha}{4\epsilon_0+\alpha^2 \mu_0}
\frac{{\rm e}^{-u z_0}-{\rm e}^{uz_0}}{1-{\rm
e}^{-2uw}\sigma}.\\
\end{split}
\label{eq:QinResult}
\end{equation}
The corresponding image charges we get are shown in
Table~\ref{tab:SlabQin}.
        
We can derive from Table~\ref{tab:SlabQin} that, when
$\epsilon_0=\epsilon$ and $\mu_0=\mu$, the expressions for the
potentials can be simplified since $\sum_{m={\rm U,L}}
q^{m+}_{k} =\sum_{m={\rm U,L}}q^{m-}_{-k}=0$, $\sum_{m={\rm
U,L}}q^{m+}_{- k}=\sum_{m={\rm U,L}}q^{m-}_{k} = q_0 (1-\sigma)
\sigma^{k-1}$, $\sum_{m={\rm U,L}}g^{m+}_{k}=\sum_{m={\rm
U,L}}g^{m-}_{-k}=q_0\frac{2\alpha}{4\epsilon_0+\alpha^2
\mu_0}\sigma^{k-1}$, and $\sum_{m={\rm U,L}}g^{m+}_{-k}=
\sum_{m={\rm U,L}}g^{m-}_{k+1}=-q_0\frac{2\alpha}{4\epsilon_0+
\alpha^2 \mu_0}\sigma^{k}$, which helps us to verify the
relation between the image charges in Table~\ref{tab:SlabQin}
and the expressions of the constants in
Eqs.~\eqref{eq:QinResult}.

\end{document}